\documentclass[onecolumn]{aastex631}
\usepackage{graphicx} 
\usepackage{color}
\usepackage{multirow}
\usepackage{hyperref}       
\usepackage{url}            
\usepackage{natbib}

\newcommand\arcsecond{{}^{\prime\prime}}

\begin{document}

\title{ 
AGNs in the extremely overdense galaxy region BOSS~1441: A Chandra observation
}

\author[0009-0003-1518-6186]{Jiahua Wu}
\affiliation{Department of Astronomy, Guangzhou University, Guangzhou 510006, China}

\author[0000-0002-4757-8622]{Liming Dou}
\affiliation{Department of Astronomy, Guangzhou University, Guangzhou 510006, China}
\correspondingauthor{Liming Dou}
\email{doulm@gzhu.edu.cn}

\author{Zheng Cai}
\affiliation{Department of Astronomy, Tsinghua University, Beijing 100084, China}

\author{Yanli Ai}
\affiliation{College of Engineering Physics, Shenzhen Technology University, Shenzhen 518118, People’s Republic of China}

\author{Shiwu Zhang}
\affiliation{Research Center for Astronomical Computing, Zhejiang Laboratory, Hangzhou 311100, China}

\author{Zhenya Zheng}
\affiliation{Shanghai Astronomical Observatory, Chinese Academy of Sciences, Shanghai 200030, China}

\author{Xiaohui Fan}
\affiliation{Steward Observatory, University of Arizona, 933 North Cherry Avenue, Rm. N204, Tucson, AZ 85721-0065, USA}

\author{Yuanyuan Su}
\affiliation{Department of Physics and Astronomy, University of Kentucky, 505 Rose Street, Lexington, KY 40506, USA}

\author{Jianfeng Wu}
\affiliation{Department of Astronomy, Xiamen University, Xiamen, Fujian 361005, China}

\begin{abstract}

We present a Chandra/ACIS-I study of X-ray sources in BOSS 1441, a protocluster at $z=2.32\pm0.02$ that exhibits a prominent overdensity of Ly$\alpha$ emitters (LAEs).
Using a 45\,ks observation, we identify seven X-ray sources spatially coincident with LAE density peaks. The average X-ray photon index for the seven sources, derived from an absorbed power-law model with Galactic absorption fixed, is 1.49 (ranging from -0.68 to 2.51), corresponding to an average luminosity of 6.85$\rm\times 10^{44} erg s^{-1}$ in the rest-frame 2-33 keV band, with individual luminosities spanning (3.57 -- 13.96)$\rm\times 10^{44} erg s^{-1}$. Three sources exhibit relatively flat spectral slopes. Two are associated with the MAMMOTH-1 nebula, while the third, located at the edge of BOSS 1441 with a $>\,5\arcmin$ offset from the LAE density peak, resides in a region with a high submillimeter-band density. We estimate the fraction of X-ray detected AGNs among the LAEs to be $11.5^{+3.8}_{-4.6}\%$, approximately double that of previously studied LAEs. This elevated fraction suggests BOSS 1441 is in a mature evolutionary stage, with even higher AGN fractions expected in massive LAEs such as PKS~1138-262. In contrast, the submillimeter galaxy population shows a lower AGN fraction ($6.9^{+6.9}_{-4.5}\%$), consistent with their typically obscured nature.
These results indicate that the protocluster's massive galaxies are evolving into the bright red sequence galaxies observed in local clusters, where AGNs likely play a critical role in quenching their star formation.

\end{abstract}


\section{Introduction}

Observational studies have shown that in the local universe, galaxy properties exhibit a clear ``environmental dependence" \citep[e.g.,][]{2009MNRAS.392.1265S}. Typically, the oldest and most evolved galaxies reside in the central regions of clusters, whereas younger, actively star-forming galaxies tend to inhabit the outer and lower-density areas. In contrast to the star formation–density relation observed in the local universe \citep{1980ApJ...236..351D,2003MNRAS.346..601G}, high-redshift galaxies in dense regions are found to experience enhanced star formation, increased interactions, and/or accelerated evolution, along with increased active galactic nuclei (AGNs) activity \citep{2007A&A...468...33E,2010ApJ...719L.126T,2011MNRAS.418..938G,2013MNRAS.428.1551K,2013ApJ...768....1M}.
Studies of X-ray AGN fraction evolution in galaxy clusters ($z \lesssim 1.3$) reveal a significant cosmic-time decline in the prevalence of AGN-hosting cluster galaxies \citep[][]{2013ApJ...768....1M,2023PASJ...75.1246H}. This observed trend underscores the importance of multi-wavelength observations targeting high-redshift overdense environments for elucidating the physical mechanisms driving nuclear activity.
Despite their scientific value, comprehensive observational campaigns in these regions remain limited.

Investigating the precursors of local massive clusters, known as protoclusters, may provide essential insights into the earliest and most active phases of cluster formation, particularly on how nuclear activity relates to the local environment.
Recent X-ray studies have revealed that AGNs are prevalent in gas-rich protoclusters, where abundant fuel reservoirs likely drive enhanced accretion \citep{Travascio2025AA,2020A&A...642A.149V,2024A&A...689A.130V}. Notably, these environments also host dusty star-forming galaxies that contribute a growing fraction to the cosmic star-formation rate density \citep{Pensabene2024AA,2018A&A...620A.202A}, suggesting AGNs play a fundamental role in regulating both galactic growth and SMBH evolution.
Protoclusters at $z \gtrsim 2$ have been systematically identified through multiple approaches. Galaxy survey-based methods reveal overdensities via spectroscopic follow-up of LBGs, as demonstrated by the $z=2.30$ HS~1700+643 \citep{Steidel2005} and $z=3.09$ SSA22 \citep{Steidel1998,Steidel2000} protoclusters. Photometric redshift techniques in wide fields (e.g., COSMOS) enable three-dimensional overdensity mapping \citep{2014ApJ...782L...3C}, while deep imaging surveys like CFHTLS identify candidates at $z\sim3-6$ through dropout selections \citep{2016ApJ...826..114T}, though all methods face limitations from finite survey volumes and spectroscopic constraints.
Alternatively, tracer-based identification targets rare signposts of massive halos: quasars \citep{Hu1996ApJ}, radio galaxies \citep{Venemans2007AA}, submillimeter galaxies (SMGs, e.g., \citealt{Chapman2004ApJ}), and Ly$\alpha$ blobs (LABs, e.g., \citealt{YangYujin2009ApJ}). While observational efficient, these tracers introduce selection biases and suffer from brief duty cycles.

A more recent approach to identifying protoclusters capitalizes on the observation that overdense regions in the early universe contain not only an abundance of dark matter and galaxies but also significant reservoirs of cold or warm dense gas. This gas can be detected via absorption against bright background continuum sources, such as quasars and galaxies. \cite{2016ApJ...833..135C} used cosmological simulations to establish a strong correlation between Ly$\alpha$ optical depth and mass overdensities on scales of $10-40 h^{-1}$ comoving Mpc (cMpc). Their findings revealed that Ly$\alpha$ 
 opacity increases with mass overdensities, particularly on scales of approximately $15~h^{-1}$~cMpc, and that regions traced by strong intergalactic medium (IGM) absorption exhibit higher overdensities than those traced by other rare tracers. These insights led to the development of the MAMMOTH (MApping the Most Massive Overdensity Through Hydrogen) technique, which leverages the extensive quasar spectral library from the SDSS-III/BOSS survey to detect rare, strong HI absorption features in the IGM and identify candidate protoclusters with exceptionally high mass concentrations.

Using the MAMMOTH technique, the massive overdensity of BOSS~1441 at $z=2.32 \pm 0.02$ is selected from the early data release of SDSS-III/BOSS \citep{2017ApJ...839..131C}. BOSS~1441 is traced by a strong IGM Ly$\alpha$ absorption group with $\tau_{eff} \geq 3.0 \times \left<\tau\right>$ within 15 $h^{-1}$ $\rm cMpc$. The absorption group is $\geq 4$ absorption systems within the projected $20 h^{-1}$ cMpc scale. Follow-up narrowband imaging and spectroscopic observations have constrained the Ly$\alpha$ emitters (LAEs) in this field. The LAE overdensity in BOSS~1441 reaches $\delta_g = 10.8 \pm 2.6$ on a 15 cMpc scale. Theoretical modeling indicates that the BOSS~1441 protocluster should collapse into a structure $z=0$ that resembles a rich local cluster with a total mass $\rm{M}_{z=0} \geq 10^{15}~\rm{M}_{\odot}$. The number of such massive clusters should be $\sim 1$ within a $10^7~\rm cMpc^3$ volume. Furthermore, BOSS~1441 is associated with an enormous Ly$\alpha$ nebula (ELAN) MAMMOTH-1 with a size of $\sim$ 442~kpc \citep{2017ApJ...837...71C}, which is used to trace the cool gas reservoir and the cosmic web. High-$\delta_g$ and the presence of the MAMMOTH-1 nebula imply that this is a gas-rich region where multiple obscured AGNs are expected to be present. 

X-ray observations are indispensable for studying AGN in dense environments, as X-ray emission can penetrate gas with high column densities. Chandra’s unparalleled spatial resolution and low background noise have been instrumental in detecting high-redshift X-ray AGNs. Over the past two decades, deep Chandra surveys have systematically uncovered AGN populations within multiple protoclusters (e.g., \citealt{Travascio2025AA,2020A&A...642A.149V,2024A&A...689A.130V,2022A&A...662A..54T,Traina2025arXiv}). Notably, \cite{2023Sci...380..494Z} identified a Compton-thick AGN (CT-AGN), designated G-2, at the core of the MAMMOTH-1 nebula through Chandra observations, underscoring the facility’s critical role in probing obscured AGN in overdense regions.

In this study, we conduct a systematic search and analysis of all X-ray-detected sources within the vicinity of the MAMMOTH-1 nebula, utilizing data obtained from Chandra observations. Additionally, we compare the AGN fraction in this region with those observed in other protoclusters to provide further insights into the environmental influence on AGN activity.
For consistency, all magnitudes in this work are reported in the AB system. We assume a standard flat $\Lambda\rm{CDM}$ cosmology with the following parameters: $H_0=70~\rm{km}~\rm{s}^{-1}~\rm{Mpc}^{-1}$, $\Omega=0.3$, and $\Omega_{\Lambda}=0.73$.

\section{Observation and Data Reduction}
\subsection{Chandra Observation \label{section:obs}}

The MAMMOTH-1 Ly$\alpha$ nebula was observed by the Chandra X-ray Observatory on 2019 April 21, using the ACIS-I camera \citep{ACIS} in the Very Faint (VFAINT) mode with a single exposure time of $\sim45~ks$ (\dataset[OBS\_ID: 20357]{https://doi.org/10.25574/20357}; PI: Z. Cai). We performed the data reduction using the CIAO software package (v4.16; \citealt{CIAO}) with the latest calibration files. We generated level 2 event files using the \texttt{chandra\_repro} script and applied fine astrometric corrections via the \texttt{fine\_astro}  script, referencing source positions from the Chandra Source Catalog (CSC 2.1; \citealt{Evans2024}).

To identify and exclude periods of high background, we examined the 2.3-7.3 keV band light curve in 100 s time bins, finding no significant flaring events ($>3\sigma$). We produced counts maps, exposure maps, and point-spread function (PSF) maps in three standard energy bands: soft (0.5-2\,keV), hard (2-7\,keV), and full (0.5-7\,keV).

We characterized the X-ray spectral hardness through the hardness ratio (HR), defined as $\rm{(H-S)/(H+S)}$, where $\rm{H}$ and $\rm{S}$ represent the background-subtracted counts in the hard and soft bands, respectively. To address low-count statistics, HRs were calculated using the Bayesian Estimation of Hardness Ratios (BEHR) method \citep{BEHR}. The resulting HR distribution spans a wide range -0.64 to 0.89 (see Table~\ref{table:objects}).

\subsection{X-ray sources detection \label{section:wavdetect}}
We performed X-ray source detection across three energy bands (soft, hard, and full) using the \texttt{WAVDETECT} algorithm \citep{WAVDETECT}. The wavelet scales were set in geometric progression (1, $\sqrt{2}$, 2, $2\sqrt{2}$, 4, $4\sqrt{2}$, 8, $8\sqrt{2}$, and 16 pixels) with a stringent false-positive probability threshold of $10^{-5}$. Initial source detection across all bands identified 142 candidate sources. Background analysis in source-free regions reveals a mean background level of $\mu < 2$ photons within a 3$\arcsec$ aperture in the full band. Poisson statistics give $\rm P(X>7|\mu=2) \approx 0.1\%$, where $\rm X$ represents photon counts within 3$\arcsec$ aperture. This indicates that net counts $>\,5$ are statistically significant ($>3\sigma$) against background fluctuations.
Applying a conservative selection criterion of $\geq 5$ net counts in at least one energy band, we obtained 81 statistically significant point sources within the ACIS-I field of view (16$\farcm$9 $\times$ 16$\farcm$9).
In the X-ray sources detection, all CSC sources except three near the field edge were detected. The average positional offset between sources near the aimpoint and their CSC counterparts is $\sim 0.13\arcsec$. Offsets increase toward the field periphery ($\sim 0.2\arcsec$--$1\arcsec$) due to larger PSF.

\begin{figure*}[hbt!]
\plotone{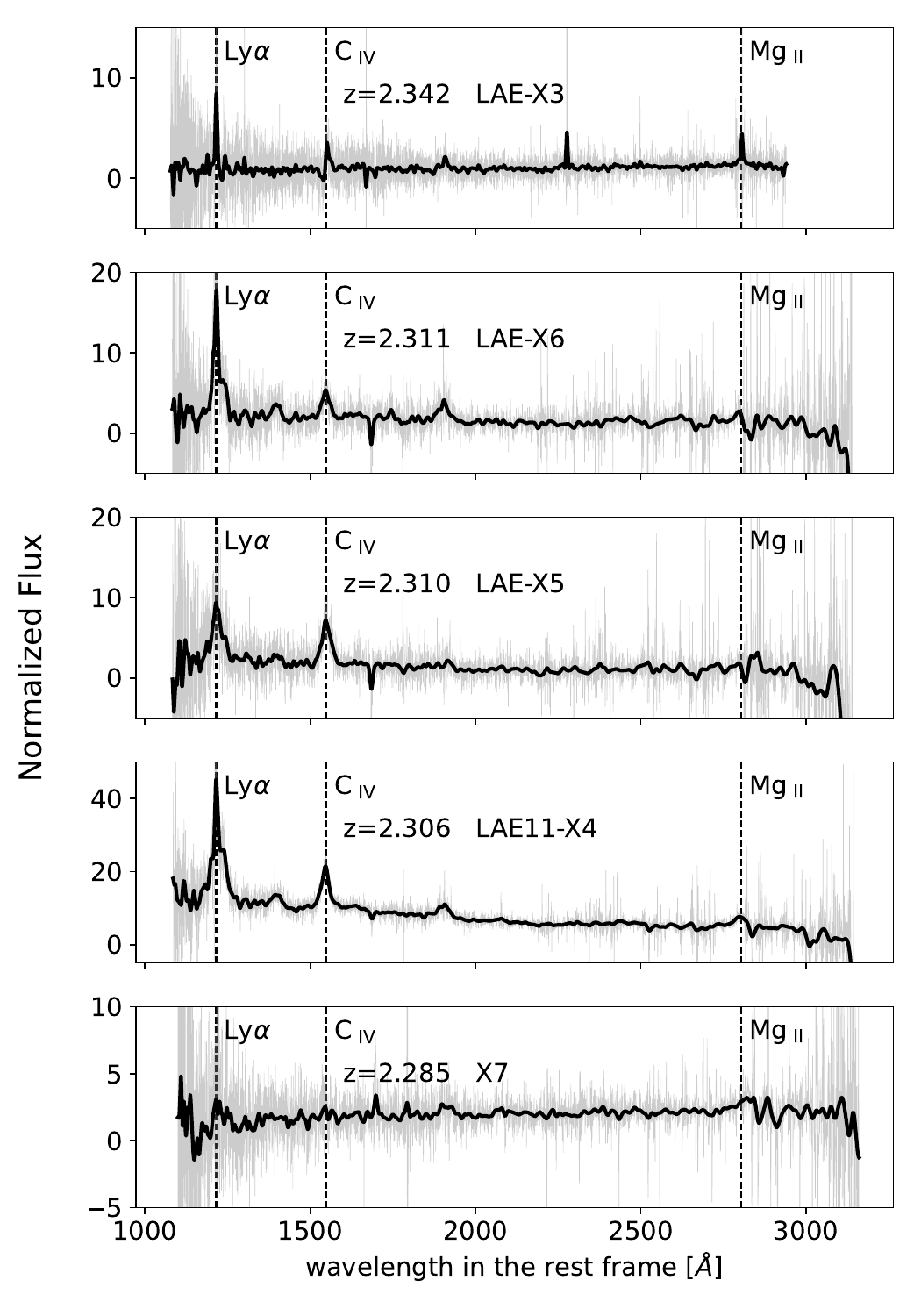}
\caption{
\label{fig:optical_spectra}
Rest-frame optical spectra of five BOSS~1441 members from SDSS/DESI observations, normalized to $10^{-17}~{\rm erg~s^{-1}~cm^{-2}~\AA^{-1}}$.
SDSS observed four sources (LAE-X5, LAE-X6, LAE11-X4, and X7), while LAE-X3 was covered in the DESI Early Data Release. All redshifts are securely confirmed through Ly$\alpha$ emission, with supplementary constraints from detected \ion{C}{4} or \ion{Mg}{2} absorption features where present.
}
\end{figure*}

We identified AGNs in the MAMMOTH-1 field through cross-matching of 81 X-ray sources with spectroscopic catalogs from SDSS \citep{2023ApJS..267...44A} and DESI Early Data Release \citep{2025arXiv250314745D}, adopting a  $3\arcsecond$ matching radius. Our analysis yielded three distinct populations: (1) 22 foreground objects at $z < 1.96$; (2) 5 confirmed protocluster members with spectroscopic redshifts $2.285 \lesssim z \lesssim 2.342$ ( Figure~\ref{fig:optical_spectra}), where redshifts were determined through Ly$\alpha$ emission lines; and (3) 54 unclassified sources requiring additional analysis.

For the unclassified sample, we incorporated optical photometry from DESI Legacy Imaging Surveys \citep{2019AJ....157..168D} and mid-infrared photometry from AllWISE \citep{2010AJ....140.1868W}. Objects with complete spectral coverage across $grz$ and $W1-W4$ bands were analyzed using \textsc{EAZY} \citep{EAZY} for photometric redshift estimation. The full multiwavelength dataset lists in Table~\ref{table:objects}. Following the methodology of \citep{2010AJ....140.1868W}, we employed WISE color-color diagnostics (their Figure 12) to effectively separate quasars from stellar and galactic contaminants. All sources in Table~\ref{table:objects} include WISE classification according to this scheme.


The MAMMOTH-1 nebula hosts populations of LAEs and SMGs, initially identified by \citet{2017ApJ...839..131C} and \citet{2018A&A...620A.202A}, respectively. Although their spatial distributions exhibit an offset which likely due to the complex gas and dust structure in BOSS 1441, we investigated potential associations with X-ray sources through catalog cross-matching.
From the \citet{2017ApJ...839..131C} LAE catalog (52 sources within the Chandra field after SDSS/DESI redshift filtering), 15 LAEs have spectroscopic confirmations (13 from MODS observations by \citet{2017ApJ...839..131C}, 2 from SDSS/DESI). Six LAEs show X-ray counterparts within a 5$\arcsec$ matching radius (LAE10-X1, LAE-X2–X6)\footnote{All counterparts except LAE10-X1 have matching radii $<2\arcsec$. The larger radius for LAE10-X1 (4.15$\arcsec$) is justified by its location within the extended MAMMOTH-1 nebula.}. Key associations include LAE10-X1 (G-2 in \citet{2023Sci...380..494Z} with CO-confirmed redshift $z=2.3116$ \citet{2019ApJ...887...86E,2021ApJ...922L..29L}) and LAE11-X4, both originally reported by \citet{2017ApJ...839..131C}. 
For the SMGs,  we combined the 850\,$\mu$m and 450\,$\mu$m catalogs from \cite{2018A&A...620A.202A}, yielding 29 sources. Only two X-ray counterparts (LAE10-X1 and LAE11-X4) were identified among 59 X-ray sources.

\begin{figure*}[hbt!]
\plotone{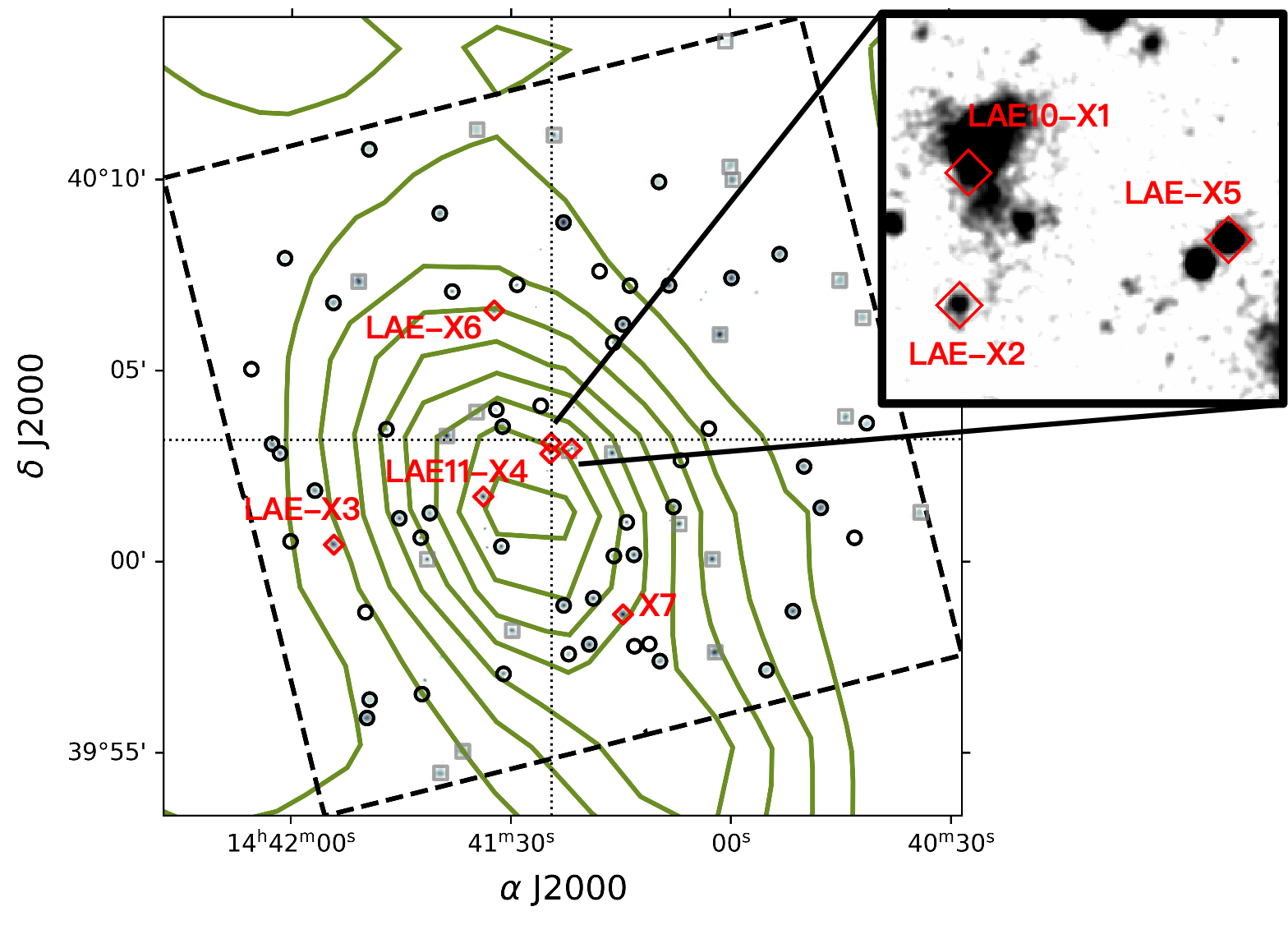}
\caption{\label{fig:coord}
Spatial distribution of X-ray-detected AGN candidates and LAE number density in the BOSS~1441 protocluster field ($z = 2.31$). The Chandra ACIS-I field of view is outlined by the black dashed square. Green contours show the LAE number density from \cite{2017ApJ...839..131C}. The Compton-thick AGN G-2 (LAE10-X1; \citealt{2023Sci...380..494Z}) is highlighted with a dotted crosshair. Symbol key: red squares = seven X-ray sources associated with BOSS~1441 (six spectroscopically confirmed; LAE-X2 membership inferred from spatial coincidence); black circles = X-ray sources without optical counterparts; gray squares = 22 spectroscopically confirmed foreground sources ($z < 2$). The inset shows the narrowband image centered on Ly$\alpha$ at $z = 2.31$, demonstrating the spatial correlation between three X-ray sources (LAE10-X1, LAE-X2, LAE-X5) and the MAMMOTH-1 Ly$\alpha$ nebula.
}
\end{figure*}

\begin{figure*}[hbt!]
\plotone{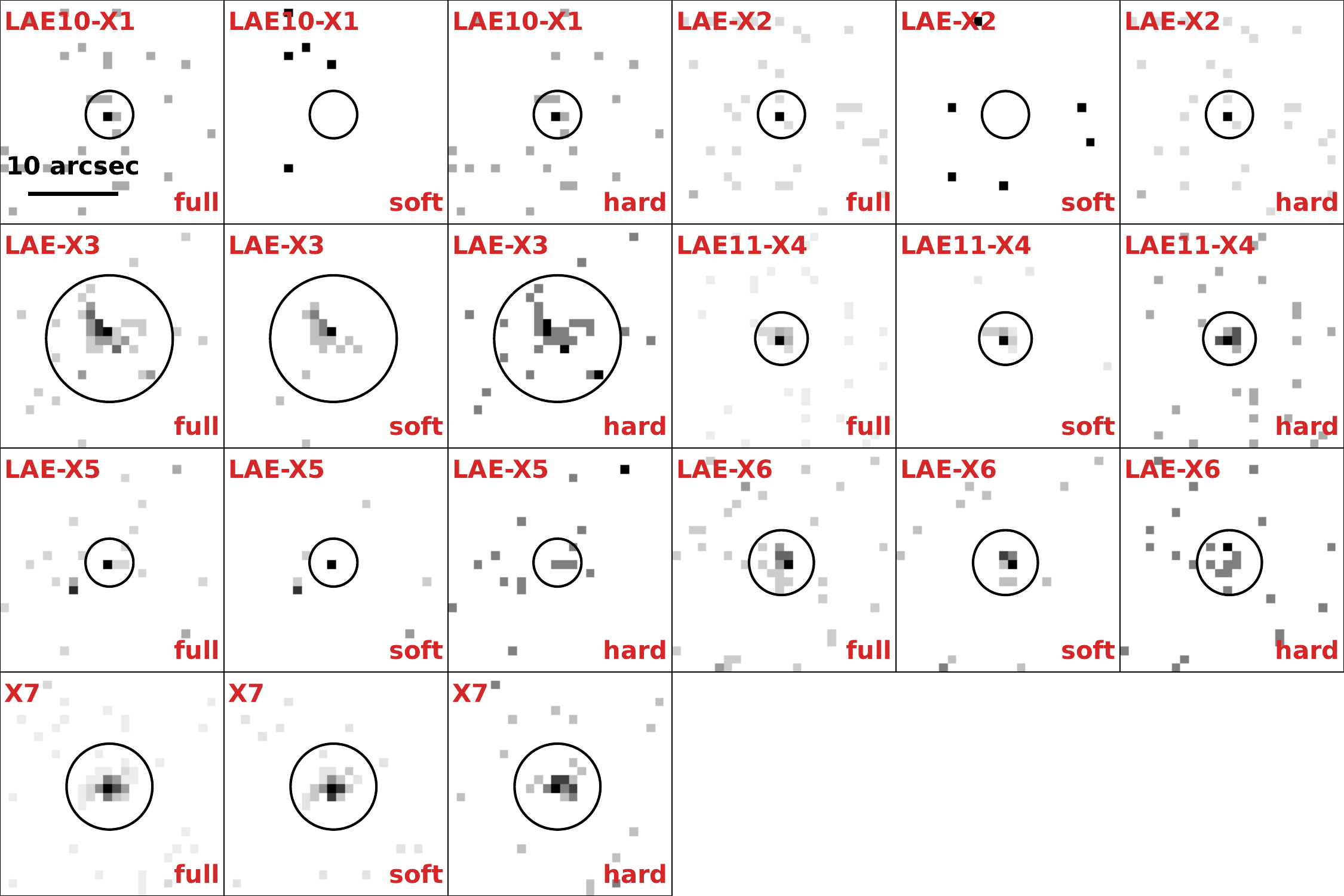}
\caption{
Chandra/ACIS $25\arcsecond \times 25\arcsecond$ cutouts of seven sources associated with BOSS~1441, displaying X-ray emission in three energy bands: full (0.5-7 keV), soft (0.5-2 keV), and hard (2-7 keV). All images are binned by a factor of two relative to the native ACIS pixel scale. The black circles indicate the 95\% encircled energy radius of the PSF at 2.3 keV, centered on each X-ray source position.
\label{fig:sources}
}
\end{figure*}

We finally classify six sources (LAE10-X1, LAE-X3, LAE11-X4, LAE-X5, LAE-X6, X7) as confirmed members of the BOSS~1441, supported by spectroscopic redshifts. For LAE-X2, although direct redshift confirmation is unavailable,  while lacking direct redshift confirmation, its spatial coincidence with both a LAE and the MAMMOTH-1 nebula (Figure~\ref{fig:coord}) suggests protocluster membership; we thus adopt $z \sim 2.31$ for LAE-X2 for subsequent analysis. The Chandra X-ray images of all confirmed members are presented in Figure~\ref{fig:sources}.

We generated source and background spectra for the seven BOSS~1441 members using the \texttt{specextract} tool, with response matrix files (RMFs) and ancillary response files (ARFs) produced concurrently. Source extraction regions were centered on each X-ray source, from circular apertures with radii ranging from 2.69$\arcsecond$ to 7.23$\arcsecond$, enclosing $\approx 95\%$ of the point spread function encircled energy at 2.3 keV.
Background spectra were extracted from concentric annuli with inner radii exceeding the source radii, except for LAE-X5, where a circular $6\arcsec$ aperture avoided contamination from a foreground source (Figure~\ref{fig:sources}). The final spectra and response files were utilized for spectral fitting and flux calculations in Section~\ref{section:xspec_analysis}.

\section{X-ray spectral analysis} \label{section:xspec_analysis}
We performed detailed X-ray spectral analysis of the seven BOSS~1441 members: six LAE counterparts and source X7. 
Spectra were binned to a minimum of two counts per bin. Spectral analysis was performed in the 0.5 to 7 keV band using the C-statistic \citep[][]{1979ApJ...228..939C, 2017A&A...605A..51K} in XSPEC\citep[][V.12.13]{XSPEC}.
The uncertainties are given at a confidence level $90\%$ for one of the interested parameters, corresponding to $\Delta C = 2.706$.
All models included Galactic absorption fixed at $\rm N_H=1.10\times 10^{20}~cm^{-2}$ \citep[``phabs" in XSPEC,][]{2016A&A...594A.116H}.

We initially modeled the X-ray spectra using a simple power-law. While all fits were statistically acceptable (Table \ref{table:fit_result_1}), the photon index for LAE10-X1 could not be well constrained ($\Gamma < 0$), yielding an upper-limit value. The remaining sources exhibited $\Gamma$ = 1.02 -- 2.51, with observed 0.5 -- 10 keV fluxes in the range  (0.85 -- 3.80)$\rm \times 10^{-14}~erg~cm^{-2}~s^{-1}$, corresponding to rest-frame 2 -- 33 keV luminosities of (3.57 -- 13.96)$\rm \times 10^{44}~erg~s^{-1}$. The extreme luminosities strongly support the identification of these sources as AGNs.

Given the flat photon indices ($\Gamma$ = -0.68 -- 1.61, $<1.8$) measured for LAE10-X1, LAE-X2, and LAE-X3, we applied an intrinsic absorbed power-law model ($zwabs*zpowerlaw$ in XSPEC) with the photon index fixed to $\Gamma=1.8$ to probe their intrinsic absorption $\rm N_H(z)$. All fits remained statistically acceptable (Table~\ref{table:fit_result_2}).
The nebula-centered source LAE10-X1 exhibits a Compton-thick (CT) AGN signature
 with a well-constrained $\rm N_H(z)=2.09^{+3.15}_{-1.69}\times 10^{24}~cm^{-2}$. This result is consistent with the CT-AGN candidate G-2 identified by \cite{2023Sci...380..494Z}, further confirming its heavily obscured nature. For the remaining six sources in our sample (LAE-X2, LAE-X3, LAE11-X4, LAE-X5, LAE-X6, and X7), we were only able to determine upper limits for their $\rm N_H(z)$ values.

For sources with identical degrees of freedom ($\nu$), the 
$C$-statistic yields no statistically significant preference between the obscured and unobscured spectral models, indicating that both provide acceptable fits to the data. To further constrain the spectral properties of low-count sources (LAE10-X1, LAE-X2, and LAE-X5), we derived the ``effective'' photon index ($\Gamma$) and intrinsic column density ($\rm N_H(z)$) from HRs, following \cite{Traina2025arXiv}. We obtain $\Gamma < 1.32$ or $\rm N_H(z) > 1.45\times 10^{23}~cm^{-2}$ for LAE10-X1 and LAE-X2; and $\Gamma > 2.3$ or $\rm N_H(z) < 2.0\times 10^{22}~cm^{-2}$ for LAE-X5, which are consistent with the $C$-statistic spectral fitting.

\section{X-ray number counts} \label{section:xray_number_counts}
Protoclusters host enhanced AGN incidence rates relative to field environments, likely facilitated by rich gas reservoirs. To quantify the X-ray AGN overdensity, we compute cumulative number counts (log$N$-log$S$) as:
\begin{equation}
N(>S) = \sum_{S_i>S} 1/\Omega,
\end{equation}
where $S_i$ represents the flux of the i-th source and $\Omega$ is the corresponding sky coverage. All fluxes were derived using the \texttt{srcflux} tool, assuming a power-law spectrum with photon index ($\Gamma=1.7$) and Galactic absorption \citep[$\rm N_H=1.10\times 10^{20}~cm^{-2}$][]{2016A&A...594A.116H}. 

\begin{figure*}[ht!]
\plotone{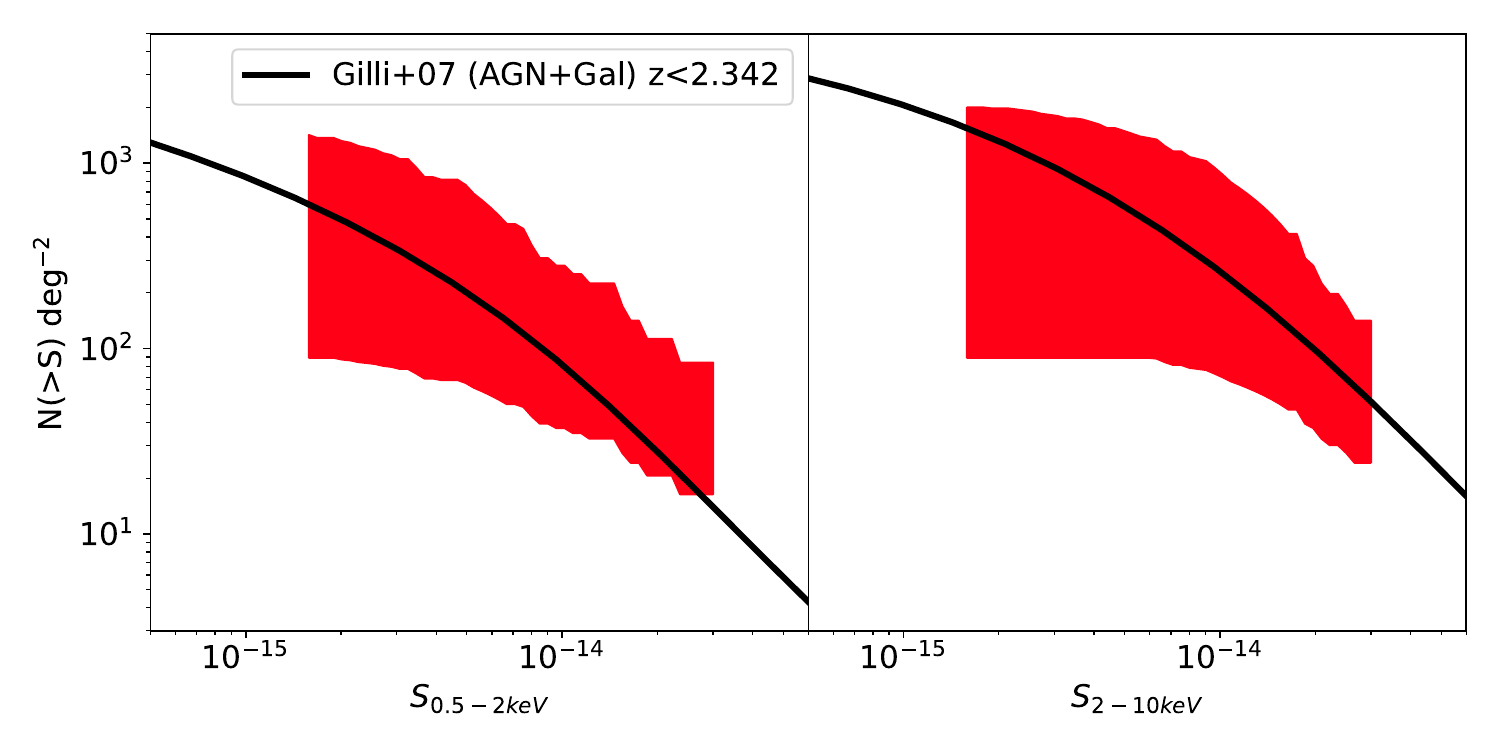}
\caption{\label{fig:xlf1} 
Cumulative number counts for X-ray sources at $z<2.342$. Left: Soft-band log$N$-log$S$ distribution within the Chandra field, compared to the composite model combining the \citet{2007A&A...463...79G} AGN population synthesis (with $z>2.7$ exponential cutoff), and \citet{2017ApJS..228....2L} galaxy counts. 
Right: Hard-band distribution with identical model comparison. Shaded regions indicate 1$\sigma$ uncertainty.
}
\end{figure*}

We first examine the line-of-sight distribution of all 81 X-ray sources.
Figure~\ref{fig:xlf1} shows the soft (0.5 -- 2~keV) and hard-band (2 -- 10~keV) number counts for sources at $z \leq 2.342$, with shaded regions indicating 1$\sigma$  (68.27\%) Poisson confidence intervals \citep{1986ApJ...303..336G}.
For comparison, we overlay predictions from: (1) the AGN population synthesis model of \citet{2007A&A...463...79G}, which includes an exponential decline at $z>2.7$, and (2) galaxy counts from the 7\,Ms Chandra Deep Field-South survey \citep{2017ApJS..228....2L}. Our measurements show good agreement with this composite model across both energy bands.

\begin{figure*}[ht!]
\plotone{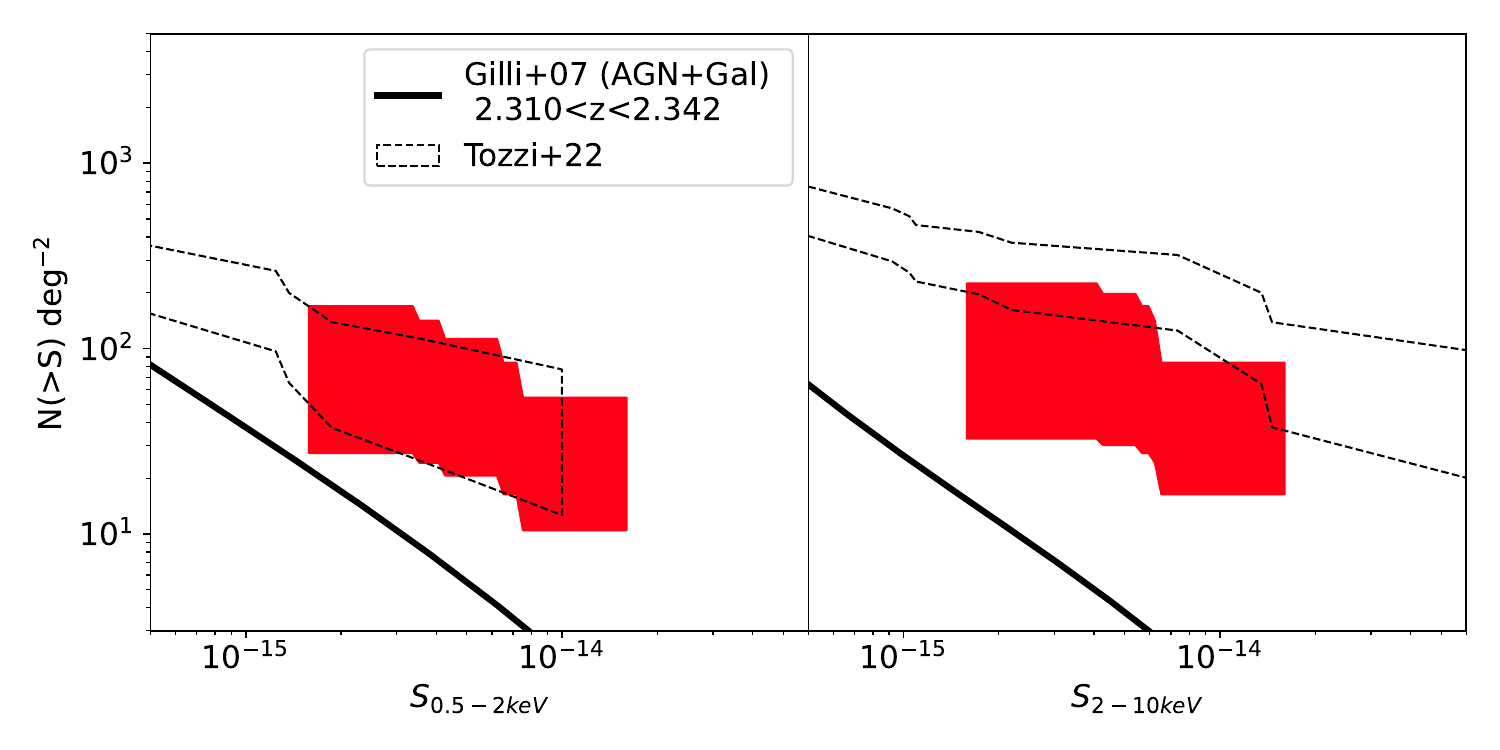}
\caption{\label{fig:xlf2} Cumulative number counts within the protocluster redshift range ($2.310 < z < 2.342$). The plotting format follows Figure~\ref{fig:xlf1}. The observed excess at fluxes $\gtrsim 10^{-15}~\mathrm{erg~cm^{-2}~s^{-1}}$ is attributed to AGN members of the BOSS~1441 system.
The dashed line shows the cumulative number counts for the PKS~1138-262 protocluster from \cite{2022A&A...662A..54T}, included for comparison with BOSS~1441.
}
\end{figure*}

Figure~\ref{fig:xlf2} further displays the log$N$-log$S$ distribution for sources in the protocluster redshift range of $2.310 < z < 2.342$.
We find enhanced normalization and a flatter slope relative to field populations, primarily driven by sources with fluxes
$\gtrsim 10^{-15}~\mathrm{erg~cm^{-2}~s^{-1}}$, which are
predominantly identified as BOSS~1441 members. For comparison, we include PKS~1138-262, a similar protocluster at a redshift $z \sim 2.2$, using deep 700~ks Chandra data alongside our shallower BOSS~1441 observations. Both systems exhibit comparable soft-band overdensities, but PKS~1138-262 shows significantly stronger hard-band enhancement. This difference may partly arise from incomplete source detection in our shallower BOSS~1441 data.

\section{Discussion}
\subsection{X-ray properties of BOSS~1441 members}
Among the seven X-ray-detected AGNs in BOSS~1441, three sources (LAE10-X1, LAE-X2, and LAE-X3) show spectral characteristics indicative of obscuration, with flat photon indices ($< 1.8$) and elevated hardness ratios. LAE10-X1, the most extreme case, is confirmed as a Compton-thick (CT) AGN through direct column density measurement. Applying the standard gas-to-dust ratio ($N_{\rm H}/A_V \approx 1.87 \times 10^{21}$ cm$^{-2}$ mag$^{-1}$; \cite{1978ApJ...224..132B}), we derive extreme extinction values: $A_V \approx 1119$ mag for LAE10-X1, with upper limits of $A_V < 840$ mag for LAE-X2 and $A_V < 60$ mag for LAE-X3. These obscuration estimates are consistent with DESI photometric observations, where LAE10-X1 (g = 23.26 mag) and LAE-X3 (g = 23.09 mag) appear as faint optical counterparts, while LAE-X2 remains undetected optically. The high fraction of obscured AGNs in this sample supports theoretical models of gas-rich, dense environments \citep{2019A&A...632A..26G,2020A&A...642A.149V}. Particularly noteworthy is the spatial coincidence of two heavily obscured AGNs (LAE10-X1 and LAE-X2, both with $\rm HR > 0.8$) with the MAMMOTH-1 nebula, a known reservoir of massive gas content.

LAE-X3 displays a flat spectral slope ($\Gamma=1.61^{+0.58}_{-0.53}$), despite its offset from the LAE density peak. Intriguingly, SCUBA-2/JCMT observations place this source within a region of elevated submillimeter source density (see Fig. 10 in \citealt{2018A&A...620A.202A}), indicating significant obscuration by both gas and dust components. This spatial configuration reveals the complex and inhomogeneous matter distribution within the BOSS~1441 system.

The spatial distribution of AGNs in BOSS~1441 reveals an interesting pattern: two unobscured AGNs (LAE11-X4 and LAE-X5) reside near the LAE density peak, with LAE11-X4 additionally exhibiting an SMG counterpart. Their redshift measurements (z=$2.3065\pm0.0005$ and z=$2.3102\pm0.0011$), combined with their projected positions, suggests these sources lie in the protocluster's periphery. Their lower obscuration is naturally explained by the reduced gas densities characteristic of such outer environments.

In addition, the presence of multiple AGNs is a well-documented feature of enormous Ly$\alpha$ nebulae (ELANe) at $z \sim 2$, as demonstrated by \citet{2014Natur.506...63C} and \citet{2015Sci...348..779H}. In MAMMOTH-1, we observe three X-ray sources spatially coinciding with Ly$\alpha$ surface brightness peaks (Figure~\ref{fig:coord} inset; \citealt{2017ApJ...837...71C}), indicating that AGN activity significantly impacts Ly$\alpha$ morphology. This interaction may operate through two primary mechanisms: (1) enhanced UV radiation from AGNs potentially amplifying Ly$\alpha$ emission, and (2) a denser circumgalactic medium facilitating intensified fluorescence via scattering processes \citep{2018ApJ...861L...3C}.


\subsection{X-ray AGN Fraction}

\begin{figure*}[ht!]
\includegraphics[width=\linewidth]{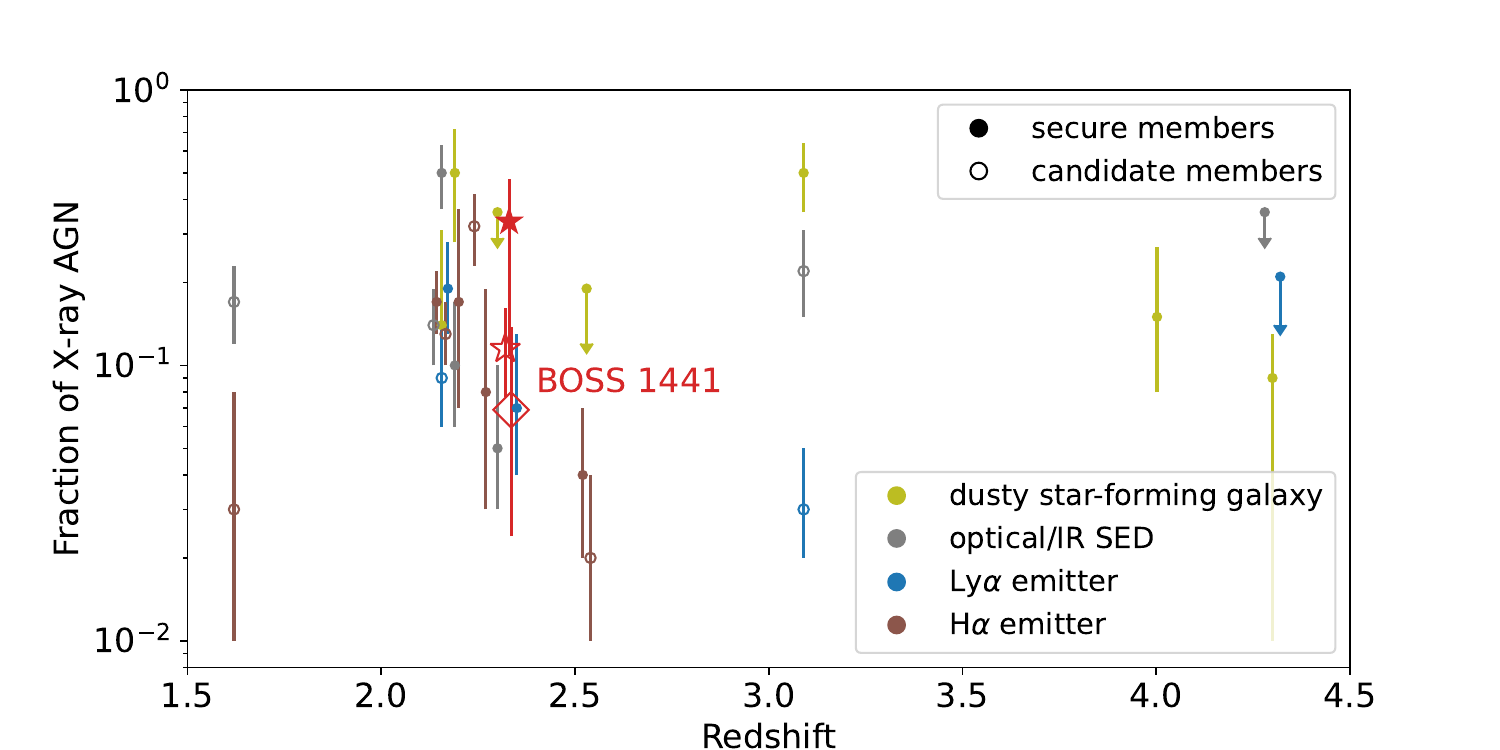}
\caption{\label{fig:result}
X-ray AGN fractions across protoclusters at $1.5 \lesssim z \lesssim 4.5$. Red stars denote AGN fractions for LAEs in BOSS~1441 (this work), while the red diamond represents SMGs in the same structure. Circles indicate AGN fractions from published protocluster studies \citep{2024A&A...689A.130V}, computed using heterogeneous galaxy tracers. Filled symbols designate measurements using spectroscopically confirmed members when available; open symbols correspond to results based on photometric member candidates.
}
\end{figure*}

The environmental dependence of AGN fraction provides critical insight into SMBH fueling mechanisms. X-ray studies have quantified this relationship across diverse environments (e.g.,  \citeauthor{2009ApJ...691..687L} \citeyear{2009ApJ...691..687L}; \citeauthor{2013ApJ...765...87L} \citeyear{2013ApJ...765...87L}; \citeauthor{10.1111/j.1365-2966.2010.16977.x} \citeyear{10.1111/j.1365-2966.2010.16977.x}; \citeauthor{2019ApJ...874...54M} \citeyear{2019ApJ...874...54M}; \citeauthor{2021A&A...654A.121P} \citeyear{2021A&A...654A.121P}). For LAEs and SMGs in BOSS~1441, we define the X-ray AGN fraction as: 
\begin{equation}
f = \frac{N_{\rm AGN}}{N_{\rm LAE/SMG}},
\end{equation}
with Poissonian uncertainties computed as double-sided 68.27\% (1$\sigma$) confidence intervals following \cite{1986ApJ...303..336G}. Our measurements yield: 
 $f_{\rm LAE} = 11.5^{+3.8}_{-4.6}\%$ for all 52 LAEs;  $f_{\rm LAE} = 33.3^{+22.5}_{-14.4}\%$ for the 15 spectroscopically confirmed LAEs; and $f_{\rm SMG} = 6.9^{+6.9}_{-4.5}\%$ for 29 SMGs.
Figure~\ref{fig:result} compares these values with other protoclusters covered by sensitive X-ray observations \citep{2024A&A...689A.130V}.
Notably, $f_{\rm LAE}$ represents one of the highest AGN fractions observed in protoclusters, while $f_{\rm SMG}$ is significantly lower. This dichotomy reflects intrinsic differences in AGN incidence between galaxy populations. Within SMGs, typical obscuration and low AGN luminosities limit detectability to only the brightest sources (\citealt{2024A&A...689A.130V}, Fig.~4).

This AGN fraction for LAEs ($f_{\rm LAE} = 11.5^{+3.8}_{-4.6}\%$) represents a factor of $\sim$2 increase over the SSA22 protocluster\citep[]{2009ApJ...691..687L}, and is elevated by 17.9$\times$ relative to the field. Fisher's Exact Test confirms this enhancement is statistically significant ($>99\%$ confidence). These results contrast with measurements from the protocluster HS~1700+64 \citep[][]{10.1111/j.1365-2966.2010.16977.x}, where a similar LAE selection yielded an AGN fraction comparable to SSA22, but with only 87\% confidence. 
The extremely high AGN activity in BOSS~1441 suggests distinct environmental conditions may be driving enhanced SMBH growth. This is further supported by (Figure~\ref{fig:xlf2}), demonstrating that the high X-ray source density cannot be solely attributed to elevated galaxy number density.

The MAMMOTH technique has identified three massive overdensities: BOSS~1441, BOSS~1244, and BOSS~1542. As characterized by \citet{2021ApJ...915...32S}, these structures exhibit distinct evolutionary states: BOSS~1441 shows signatures of advanced virialization, BOSS~1244 comprises one or two merging protoclusters, and BOSS~1542 forms a filamentary network. These morphologies align with the three-phase cluster formation framework of \citet{10.1093/mnras/sty2618}: (1) Growing phase ($z \geq 3$): large-scale structure collapse; (2) Maturing phase ($z=2-3$): transition from dusty starbursts to quenched populations; (3) Declining phase ($z \leq 2$): emergence of red sequences. 

Our Chandra observations reveal an elevated X-ray AGN fraction in BOSS~1441 ($f_{\rm LAE} =11.5^{+3.8}_{-4.6}\%$). This enhanced AGN activity provides direct observational evidence for ongoing SMBH growth, strongly supporting \citeauthor{2021ApJ...915...32S}'s virialization hypothesis. The measurement confirms the predicted correlation between structural maturity and AGN activity during the critical maturation phase at $z = 2-3$.

The dense protocluster PKS~1138-262, associated with a radio galaxy galaxy at $z=2.156$ \citep{1997ApJS..109....1C}, exhibits striking evidence for AGN-driven quenching.  \cite{10.1093/mnras/sty2618} report a stellar mass-dependent AGN fraction among H$\alpha$ emitters (HAEs), with $>60\%~(4/6)$ of massive systems ($\rm M_{\star} = 10^{11-12.5} M_{\odot}$) hosting X-ray AGNs. These HAEs occupy the post-starburst region in rest-frame \textit{UVJ} color space, displaying intermediate properties between star-forming galaxies and passive populations, suggesting AGN feedback may drive their transition from star-forming to quiescent states.

For BOSS~1441, a protocluster in the maturing phase with ongoing galaxy quenching, we similarly predict elevated AGN fractions among massive LAEs. This aligns with findings in SSA22, where \citet{2009ApJ...691..687L} demonstrated that LAEs/LBGs with stellar masses 1.2\textendash1.8× higher than field galaxies host proportionally enhanced AGN incidence. Together, these results underscore SMBH-galaxy co-evolution signatures during cluster assembly.

For five SDSS/DESI AGNs (LAE-X3, LAE11-X4, LAE-X5, LAE-X6, and X7), we estimate the virial SMBH masses via \ion{C}{4} emission:
\begin{equation}
\log\left(\frac{M_{\rm{BH}}}{M_\odot}\right) = a + b\log\left(\frac{\lambda L_\lambda}{10^{44}~\rm{erg~s^{-1}}}\right) + 2\log\left(\frac{\rm{FWHM}}{\rm{km~s^{-1}}}\right),
\end{equation}
where $a=0.66$ and $b=0.53$ are empirical coefficients for \ion{C}{4} \citep{2022ApJS..263...42W}.
The calculated values $M_{\rm{BH}}$ = (2.24 -- 8.32)$\times 10^8~M_\odot$ align with the DR16Q sample average \citep{2022ApJS..263...42W}, suggesting that environmental mechanisms, rather than individual SMBH growth processes, dominate the elevated AGN fraction in BOSS~1441.
This apparent contradiction may arise from obscured AGNs in gas-rich environments: our X-ray spectral analysis (Section~\ref{section:xspec_analysis}) reveals several Compton-thick candidates with intrinsic $L_{\text{X}}\,>\,10^{44}\,erg\,s^{-1}$, where dense circumnuclear material could simultaneously obscure AGNs and fuel rapid SMBH growth.

\section{Conclusion}
Our \textit{Chandra} X-ray study of the BOSS~1441 protocluster, host to the MAMMOTH-1 nebula, reveals a distinct population of AGNs embedded in its gas-rich, overdense environment. The main findings are summarized as follow:

\begin{enumerate}
    \item The MAMMOTH-1 neblua is associated with three X-ray AGNs (LAE10-X1, LAE-X2, LAE-X5), indicating that its Ly$\alpha$ luminosity is powered by multiple AGNs rather than a single ionizing source.
    \item X-ray hardness ratios and spectral modeling identify heavily obscured AGNs, including a Compton-thick AGN with $\rm N_H(z)$ = $\rm {2.09^{+3.15}_{-1.69}} \times 10^{24}~\rm cm^{-2}$. This obscuration aligns with the protocluster's gas-rich conditions, where dense material both shrouds AGNs and fuels rapid supermassive black hole growth.
    \item An obscured object (LAE-X3) is spatially offset from the LAE density peak yet coincides with a submillimeter-band overdensity. This configuration reveals concurrent gas/dust obscuration and traces inhomogeneous matter distribution within the protocluster.
    \item The X-ray AGN fraction in BOSS~1441 is $11.5^{+3.8}_{-4.6}\%$ for the full LAE sample (52 sources), representing a statistically significant enhancement ($>99\%$ confidence). For the subset of 15 spectroscopically confirmed LAEs, this fraction rises to $33.3^{+22.5}_{-14.4}\%$, ranking among the highest values observed in protoclusters. Both fractions rank among the highest observed in protoclusters. These elevated fractions support BOSS~1441's classification as a system in a maturing phase of virialization, contrasting with younger protoclusters (e.g., SSA22) and mirroring evolved counterparts (e.g., PKS~1138-262) where AGN feedback likely drives galaxy quenching. These signatures indicate BOSS~1441 is transitioning to a phase of declining star formation, with massive members evolving into red-sequence galaxies.
\end{enumerate}

Collectively, these findings establish gas-rich protoclusters as pivotal laboratories for SMBH-galaxy co-evolution, where environmental fueling mechanisms govern the emergence of massive quiescent galaxies in the early universe.

\begin{longrotatetable}
\begin{deluxetable}{cccccccccccccccccc}
\tabletypesize{\tiny}
\tablecaption{X-ray Source Catalog in the Chandra Field of View} \label{table:objects}
\tablehead{
    \colhead{(1)} &
    \colhead{(2)} &
    \colhead{(3)} &
    \colhead{(4)} &
    \colhead{(5)} &
    \colhead{(6)} &
    \colhead{(7)} &
    \colhead{(8)} &
    \colhead{(9)} &
    \colhead{(10)} &
    \colhead{(11)} &
    \colhead{(12)} &
    \colhead{(13)} &
    \colhead{(14)} &
    \colhead{(15)} &
    \colhead{(16)} &
    \colhead{(17)} &
    \colhead{(18)} \\
    \colhead{$XID$} &
    \colhead{${\alpha}_{\rm X}$} &
    \colhead{${\beta}_{\rm X}$} &
    \colhead{$R_{95}$} &
    \colhead{z$_{\rm sp}$} &
    \colhead{z$_{\rm ph}$} &
    \colhead{$Net cts$} &
    \colhead{$S_{0.5-2~keV}$} &
    \colhead{$S_{2-10~keV}$} &
    \colhead{$HR$} &
    \colhead{$g$} &
    \colhead{$r$} &
    \colhead{$z$} &
    \colhead{$W1$} &
    \colhead{$W2$} &
    \colhead{$W3$} &
    \colhead{$W4$} &
    \colhead{Type}
}
\startdata
LAE10-X1 & 220.3520 & 40.0520 & 2.71$\arcsecond$ & 2.312 & $2.210$ & $5.96  ^{+3.00} _{-2.40}$  & $<0.03$ & $0.42^{+0.20}_{-0.16}$                & $0.86 ^{+0.14}_{-0.01}$ & 23.26 & 23.03 & 23.11 & 19.45 & 19.07 & 17.68 & 15.40 & QSO \\
LAE-X2 & 220.3525 & 40.0473 & 2.69$\arcsecond$ & --- & ---         & $7.80  ^{+3.24} _{-2.90}$  & $<0.03$ & $0.65^{+0.25}_{-0.21}$                & $0.89 ^{+0.11}_{-0.01}$ & --- & --- & --- & --- & --- & --- & --- & --- \\
LAE-X3 & 220.4756 & 40.0074 & 7.23$\arcsecond$ & 2.342 & $1.284$   & $44.00 ^{+7.40} _{-7.08}$  & $0.74^{+0.19}_{-0.17}$ & $1.86^{+0.48}_{-0.44}$ & $0.09 ^{+0.16}_{-0.16}$ & 23.09 & 22.40 & 21.30 & 20.04 & 20.22 & 17.61 & 16.07 & ULIRG; LINER; Starburst \\
LAE11-X4 & 220.3906 & 40.0284 & 3.00$\arcsecond$ & 2.306 & $1.300$ & $30.00 ^{+5.86} _{-5.53}$  & $0.63^{+0.15}_{-0.13}$ & $0.61^{+0.26}_{-0.22}$ & $-0.37^{+0.15}_{-0.21}$ & 20.51 & 20.53 & 20.26 & 20.93 & 20.03 & 18.07 & 15.74 & ULIRG/LINER; Obscured AGN \\
LAE-X5 & 220.3403 & 40.0496 & 2.74$\arcsecond$ & 2.310 & ---       & $8.35  ^{+3.32} _{-2.76}$  & $0.41^{+0.22}_{-0.17}$ & $0.56^{+0.44}_{-0.31}$ & $-0.16^{+0.35}_{-0.35}$ & 21.74 & 21.79 & 21.52 & --- & --- & --- & --- & --- \\
LAE-X6 & 220.3845 & 40.1097 & 3.70$\arcsecond$ & 2.311 & $1.396$   & $20.05 ^{+4.93} _{-4.49}$  & $0.34^{+0.12}_{-0.10}$ & $0.64^{+0.25}_{-0.21}$ & $-0.11^{+0.22}_{-0.24}$ & 22.11 & 21.87 & 21.61 & 20.59 & 20.68 & 17.65 & 16.11 & ULIRG; LINER; Starburst \\
X7 & 220.3111 & 39.9771 & 4.91$\arcsecond$ & 2.285 & $1.293$       & $70.80 ^{+8.71} _{-8.62}$  & $1.79^{+0.26}_{-0.26}$ & $1.62^{+0.42}_{-0.38}$ & $-0.37^{+0.10}_{-0.13}$ & 22.67 & 21.75 & 20.74 & 19.63 & 19.41 & 18.03 & 15.34 & QSO \\
X8 & 220.2492 & 40.1238 & 6.58$\arcsecond$ & --- & $3.263$         & $56.21 ^{+8.17} _{-8.09}$  & $0.96^{+0.20}_{-0.18}$ & $1.82^{+0.45}_{-0.41}$ & $-0.04^{+0.13}_{-0.16}$ & 22.38 & 21.61 & 21.16 & 19.04 & 18.89 & 18.09 & 15.67 & QSO \\
X9 & 220.3163 & 40.0026 & 3.58$\arcsecond$ & --- & ---             & $9.20  ^{+3.39} _{-3.05}$  & $0.02^{+0.05}_{-0.02}$ & $0.48^{+0.23}_{-0.19}$ & $0.77 ^{+0.23}_{-0.05}$ & 24.15 & 23.54 & 22.09 & --- & --- & --- & --- & --- \\
X10 & 220.3450 & 40.1481 & 5.58$\arcsecond$ & --- & $3.404$        & $172.41^{+13.38}_{-13.25}$ & $3.47^{+0.34}_{-0.34}$ & $4.86^{+0.62}_{-0.61}$ & $-0.21^{+0.07}_{-0.08}$ & 24.47 & 23.09 & 21.65 & 18.90 & 18.42 & 18.10 & 15.37 & QSO \\
X11 & 220.3803 & 40.0068 & 3.29$\arcsecond$ & --- & ---            & $12.00 ^{+3.98} _{-3.65}$  & $0.22^{+0.10}_{-0.08}$ & $0.18^{+0.18}_{-0.14}$ & $-0.22^{+0.30}_{-0.33}$ & 24.01 & 23.13 & 22.60 & --- & --- & --- & --- & --- \\
X12 & 220.3111 & 40.1037 & 3.56$\arcsecond$ & --- & $2.108$        & $42.01 ^{+6.83} _{-6.49}$  & $0.81^{+0.17}_{-0.15}$ & $0.97^{+0.31}_{-0.26}$ & $-0.27^{+0.13}_{-0.18}$ & 20.67 & 20.55 & 20.79 & 19.25 & 18.84 & 17.72 & 16.04 & QSO \\
X13 & 220.3797 & 40.0591 & 2.91$\arcsecond$ & --- & $1.254$        & $14.80 ^{+4.05} _{-3.75}$  & $0.33^{+0.17}_{-0.13}$ & $1.23^{+0.45}_{-0.36}$ & $0.21 ^{+0.23}_{-0.27}$ & 20.87 & 20.18 & 19.84 & 19.35 & 20.11 & 18.29 & 16.12 & ULIRG; LINER; Starburst \\
X14 & 220.5109 & 40.0515 & 8.72$\arcsecond$ & --- & ---            & $25.18 ^{+5.95} _{-5.58}$  & $0.37^{+0.15}_{-0.13}$ & $1.02^{+0.43}_{-0.38}$ & $0.31 ^{+0.22}_{-0.20}$ & 24.87 & 23.12 & 21.85 & --- & --- & --- & --- & --- \\
X15 & 220.2824 & 40.0239 & 3.70$\arcsecond$ & --- & $1.270$        & $15.80 ^{+4.47} _{-4.16}$  & $0.29^{+0.11}_{-0.09}$ & $0.39^{+0.24}_{-0.20}$ & $-0.21^{+0.27}_{-0.27}$ & 24.15 & 23.29 & 21.54 & 19.68 & 19.98 & 17.73 & 15.67 & Seyfert galaxy \\
X16 & 220.3280 & 39.9841 & 4.34$\arcsecond$ & --- & $1.671$        & $16.01 ^{+4.91} _{-4.51}$  & $0.35^{+0.13}_{-0.11}$ & $0.38^{+0.26}_{-0.22}$ & $-0.37^{+0.32}_{-0.30}$ & 23.01 & 22.49 & 22.19 & 20.45 & 19.80 & 17.75 & 16.01 & Seyfert galaxy \\
X17 & 220.2145 & 39.9785 & 9.57$\arcsecond$ & --- & $4.329$        & $38.20 ^{+7.39} _{-7.09}$  & $0.76^{+0.21}_{-0.18}$ & $1.76^{+0.53}_{-0.48}$ & $0.01 ^{+0.19}_{-0.18}$ & 23.98 & 22.34 & 20.88 & 19.37 & 19.26 & 18.18 & 16.00 & QSO \\
X18 & 220.3050 & 40.0031 & 3.78$\arcsecond$ & --- & $1.244$        & $22.00 ^{+5.07} _{-4.81}$  & $0.11^{+0.08}_{-0.06}$ & $1.33^{+0.35}_{-0.31}$ & $0.66 ^{+0.21}_{-0.12}$ & 22.62 & 21.57 & 20.17 & 19.25 & 19.80 & 17.60 & 15.95 & ULIRG; LINER; Starburst \\
X19 & 220.2848 & 40.1207 & 4.65$\arcsecond$ & --- & $1.432$        & $37.21 ^{+6.60} _{-6.32}$  & $0.63^{+0.16}_{-0.14}$ & $1.28^{+0.36}_{-0.31}$ & $-0.09^{+0.19}_{-0.16}$ & 21.44 & 21.14 & 20.97 & 20.13 & 19.87 & 18.14 & 15.89 & Seyfert galaxy \\
X20 & 220.3303 & 39.9641 & 5.82$\arcsecond$ & --- & $1.310$        & $33.20 ^{+6.37} _{-6.11}$  & $0.60^{+0.17}_{-0.15}$ & $1.18^{+0.38}_{-0.33}$ & $-0.00^{+0.20}_{-0.18}$ & 23.99 & 22.72 & 20.87 & 18.66 & 18.03 & 16.51 & 15.55 & QSO \\
X21 & 220.3449 & 39.9810 & 4.40$\arcsecond$ & --- & $1.368$        & $20.40 ^{+5.01} _{-4.73}$  & $0.21^{+0.10}_{-0.08}$ & $0.91^{+0.33}_{-0.28}$ & $0.36 ^{+0.24}_{-0.19}$ & 21.85 & 21.38 & 20.51 & 19.13 & 19.06 & 18.16 & 15.31 & QSO \\
X22 & 220.2902 & 39.9568 & 7.45$\arcsecond$ & --- & ---            & $12.61 ^{+4.96} _{-4.61}$  & $0.15^{+0.13}_{-0.10}$ & $0.82^{+0.41}_{-0.36}$ & $0.44 ^{+0.56}_{-0.15}$ & 24.78 & 23.54 & 22.16 & --- & --- & --- & --- & --- \\
X23 & 220.3791 & 39.9512 & 7.25$\arcsecond$ & --- & ---            & $10.61 ^{+4.62} _{-4.25}$  & $0.27^{+0.13}_{-0.11}$ & $0.31^{+0.28}_{-0.24}$ & $-0.35^{+0.19}_{-0.65}$ & 25.04 & 23.92 & 22.47 & --- & --- & --- & --- & --- \\
X24 & 220.4155 & 40.1521 & 7.10$\arcsecond$ & --- & ---            & $14.00 ^{+4.93} _{-4.59}$  & $0.34^{+0.15}_{-0.13}$ & $0.80^{+0.40}_{-0.35}$ & $-0.22^{+0.42}_{-0.33}$ & --- & --- & --- & 20.40 & 19.86 & 18.01 & 16.07 & Seyfert galaxy \\
X25 & 220.4761 & 40.1130 & 7.52$\arcsecond$ & --- & $1.278$        & $17.80 ^{+5.21} _{-4.89}$  & $0.52^{+0.17}_{-0.14}$ & $0.15^{+0.27}_{-0.15}$ & $-0.52^{+0.29}_{-0.31}$ & 23.61 & 22.85 & 21.25 & 19.64 & 19.62 & 17.96 & 15.97 & Seyfert galaxy \\
X26 & 220.4567 & 39.9319 & 12.43$\arcsecond$ & --- & ---           & $60.28 ^{+8.96} _{-8.87}$  & $0.81^{+0.24}_{-0.21}$ & $3.50^{+0.75}_{-0.69}$ & $0.43 ^{+0.15}_{-0.12}$ & 24.42 & 23.57 & 22.21 & --- & --- & --- & --- & --- \\
X27 & 220.2218 & 40.1342 & 8.77$\arcsecond$ & --- & ---            & $20.00 ^{+5.93} _{-5.60}$  & $0.29^{+0.14}_{-0.12}$ & $0.90^{+0.40}_{-0.36}$ & $0.19 ^{+0.35}_{-0.24}$ & 24.44 & 23.35 & 24.02 & --- & --- & --- & --- & --- \\
X28 & 220.2080 & 40.0415 & 7.34$\arcsecond$ & --- & ---            & $19.60 ^{+5.42} _{-5.03}$  & $0.25^{+0.12}_{-0.09}$ & $0.76^{+0.36}_{-0.32}$ & $0.23 ^{+0.28}_{-0.24}$ & 23.30 & 22.49 & 21.45 & --- & --- & --- & --- & --- \\
X29 & 220.4865 & 40.0310 & 7.17$\arcsecond$ & --- & ---            & $16.80 ^{+4.88} _{-4.51}$  & $0.20^{+0.12}_{-0.09}$ & $0.91^{+0.36}_{-0.31}$ & $0.36 ^{+0.28}_{-0.26}$ & 24.04 & 23.49 & 23.06 & --- & --- & --- & --- & --- \\
X30 & 220.1985 & 40.0234 & 8.43$\arcsecond$ & --- & ---            & $44.61 ^{+7.83} _{-7.43}$  & $0.78^{+0.20}_{-0.17}$ & $2.46^{+0.53}_{-0.49}$ & $0.01 ^{+0.17}_{-0.17}$ & 22.29 & 21.53 & 21.16 & --- & --- & --- & --- & --- \\
X31 & 220.5228 & 40.0840 & 10.24$\arcsecond$ & --- & ---           & $11.80 ^{+5.68} _{-5.23}$  & $0.30^{+0.18}_{-0.15}$ & $0.68^{+0.44}_{-0.40}$ & $-0.16^{+0.26}_{-0.84}$ & 24.41 & 24.37 & 23.46 & --- & --- & --- & --- & --- \\
X32 & 220.4558 & 40.1799 & 11.78$\arcsecond$ & --- & $1.398$       & $18.41 ^{+6.27} _{-5.92}$  & $0.12^{+0.14}_{-0.12}$ & $1.64^{+0.61}_{-0.56}$ & $0.76 ^{+0.24}_{-0.06}$ & 24.92 & 23.85 & 22.64 & 19.42 & 20.01 & 18.19 & 15.52 & ULIRG; LINER; Starburst \\
X33 & 220.3090 & 40.0172 & 3.14$\arcsecond$ & --- & ---            & $10.00 ^{+3.55} _{-3.21}$  & $0.20^{+0.09}_{-0.07}$ & $0.23^{+0.17}_{-0.13}$ & $-0.32^{+0.33}_{-0.33}$ & --- & --- & --- & --- & --- & --- & --- & --- \\
X34 & 220.1723 & 40.0605 & 10.31$\arcsecond$ & --- & ---           & $11.61 ^{+5.36} _{-4.96}$  & $0.17^{+0.14}_{-0.11}$ & $0.67^{+0.45}_{-0.41}$ & $0.34 ^{+0.66}_{-0.17}$ & --- & --- & --- & --- & --- & --- & --- & --- \\
X35 & 220.5003 & 40.0088 & 8.90$\arcsecond$ & --- & ---            & $9.40  ^{+4.50} _{-4.11}$  & $0.36^{+0.16}_{-0.13}$ & $0.18^{+0.27}_{-0.18}$ & $-0.64^{+0.11}_{-0.36}$ & --- & --- & --- & --- & --- & --- & --- & --- \\
X36 & 220.3579 & 40.0682 & 2.91$\arcsecond$ & --- & ---            & $5.77  ^{+2.64} _{-2.38}$  & $<0.04$ & $0.39^{+0.22}_{-0.17}$                & $0.85 ^{+0.15}_{-0.01}$ & --- & --- & --- & --- & --- & --- & --- & --- \\
X37 & 220.4458 & 40.0579 & 4.28$\arcsecond$ & --- & ---            & $7.60  ^{+3.24} _{-2.91}$  & $0.14^{+0.09}_{-0.07}$ & $0.29^{+0.23}_{-0.18}$ & $-0.07^{+0.49}_{-0.36}$ & --- & --- & --- & --- & --- & --- & --- & --- \\
X38 & 220.3244 & 40.1268 & 4.30$\arcsecond$ & --- & ---            & $6.40  ^{+3.12} _{-2.74}$  & $0.06^{+0.06}_{-0.04}$ & $0.27^{+0.21}_{-0.17}$ & $0.36 ^{+0.64}_{-0.14}$ & --- & --- & --- & --- & --- & --- & --- & --- \\
X39 & 220.3046 & 39.9632 & 6.41$\arcsecond$ & --- & ---            & $11.10 ^{+4.16} _{-3.81}$  & $0.01^{+0.08}_{-0.01}$ & $1.05^{+0.36}_{-0.31}$ & $0.82 ^{+0.18}_{-0.04}$ & --- & --- & --- & --- & --- & --- & --- & --- \\
X40 & 220.1793 & 40.0104 & 10.73$\arcsecond$ & --- & ---           & $14.80 ^{+5.65} _{-5.20}$  & $0.02^{+0.12}_{-0.02}$ & $1.76^{+0.56}_{-0.51}$ & $0.79 ^{+0.21}_{-0.05}$ & --- & --- & --- & --- & --- & --- & --- & --- \\
X41 & 220.4577 & 39.9780 & 7.86$\arcsecond$ & --- & ---            & $13.00 ^{+4.85} _{-4.45}$  & $0.01^{+0.09}_{-0.01}$ & $1.38^{+0.43}_{-0.38}$ & $0.83 ^{+0.17}_{-0.04}$ & --- & --- & --- & --- & --- & --- & --- & --- \\
X42 & 220.2781 & 40.0443 & 3.50$\arcsecond$ & --- & ---            & $5.40  ^{+2.93} _{-2.56}$  & $0.11^{+0.08}_{-0.06}$ & $0.25^{+0.20}_{-0.16}$ & $-0.27^{+0.22}_{-0.73}$ & --- & --- & --- & --- & --- & --- & --- & --- \\
X43 & 220.4255 & 39.9424 & 9.42$\arcsecond$ & --- & ---            & $23.32 ^{+6.17} _{-5.87}$  & $0.48^{+0.17}_{-0.14}$ & $0.68^{+0.44}_{-0.40}$ & $-0.11^{+0.29}_{-0.25}$ & --- & --- & --- & --- & --- & --- & --- & --- \\
X44 & 220.4262 & 40.0105 & 4.32$\arcsecond$ & --- & ---            & $7.80  ^{+3.44} _{-3.08}$  & $0.14^{+0.09}_{-0.07}$ & $0.22^{+0.20}_{-0.17}$ & $-0.11^{+0.44}_{-0.46}$ & --- & --- & --- & --- & --- & --- & --- & --- \\
X45 & 220.4384 & 40.0190 & 4.47$\arcsecond$ & --- & ---            & $15.00 ^{+4.49} _{-4.18}$  & $0.13^{+0.09}_{-0.06}$ & $0.68^{+0.29}_{-0.25}$ & $0.43 ^{+0.28}_{-0.24}$ & --- & --- & --- & --- & --- & --- & --- & --- \\
X46 & 220.5064 & 40.0474 & 8.41$\arcsecond$ & --- & ---            & $15.93 ^{+5.39} _{-5.06}$  & $0.16^{+0.13}_{-0.10}$ & $0.98^{+0.45}_{-0.39}$ & $0.56 ^{+0.44}_{-0.12}$ & --- & --- & --- & --- & --- & --- & --- & --- \\
X47 & 220.4211 & 40.0213 & 3.84$\arcsecond$ & --- & ---            & $6.00  ^{+3.29} _{-2.93}$  & $0.08^{+0.07}_{-0.05}$ & $0.47^{+0.27}_{-0.22}$ & $0.03 ^{+0.62}_{-0.45}$ & --- & --- & --- & --- & --- & --- & --- & --- \\
X48 & 220.3832 & 40.0665 & 2.98$\arcsecond$ & --- & ---            & $5.20  ^{+2.72} _{-2.35}$  & $0.05^{+0.07}_{-0.04}$ & $0.40^{+0.23}_{-0.18}$ & $0.37 ^{+0.63}_{-0.18}$ & --- & --- & --- & --- & --- & --- & --- & --- \\
X49 & 220.2295 & 39.9528 & 10.86$\arcsecond$ & --- & ---           & $14.20 ^{+5.39} _{-5.03}$  & $0.22^{+0.17}_{-0.13}$ & $1.00^{+0.51}_{-0.46}$ & $0.37 ^{+0.45}_{-0.29}$ & --- & --- & --- & --- & --- & --- & --- & --- \\
X50 & 220.2905 & 40.1657 & 8.05$\arcsecond$ & --- & ---            & $18.61 ^{+5.40} _{-5.09}$  & $0.24^{+0.12}_{-0.10}$ & $1.11^{+0.42}_{-0.37}$ & $0.23 ^{+0.31}_{-0.25}$ & --- & --- & --- & --- & --- & --- & --- & --- \\
X51 & 220.4552 & 39.9400 & 11.38$\arcsecond$ & --- & ---           & $27.00 ^{+6.86} _{-6.50}$  & $0.52^{+0.19}_{-0.17}$ & $1.35^{+0.53}_{-0.48}$ & $0.06 ^{+0.27}_{-0.23}$ & --- & --- & --- & --- & --- & --- & --- & --- \\
X52 & 220.3166 & 40.0956 & 3.24$\arcsecond$ & --- & ---            & $7.00  ^{+3.27} _{-2.93}$  & $0.01^{+0.05}_{-0.01}$ & $0.32^{+0.21}_{-0.17}$ & $0.75 ^{+0.25}_{-0.05}$ & 22.29 & 20.92 & 20.15 & --- & --- & --- & --- & --- \\
X53 & 220.4084 & 40.1180 & 4.42$\arcsecond$ & --- & $1.375$        & $8.51  ^{+3.63} _{-3.20}$  & $0.09^{+0.08}_{-0.06}$ & $0.32^{+0.23}_{-0.19}$ & $0.24 ^{+0.44}_{-0.35}$ & 25.21 & 24.49 & 22.54 & 19.95 & 19.63 & 18.05 & 16.09 & QSO \\
X54 & 220.2622 & 40.0582 & 4.02$\arcsecond$ & --- & ---            & $3.80  ^{+2.70} _{-2.33}$  & $<0.07$ & $0.55^{+0.28}_{-0.23}$                & $0.65 ^{+0.35}_{-0.06}$ & 24.21 & 24.10 & 22.69 & --- & --- & --- & --- & --- \\
X55 & 220.5038 & 40.1324 & 10.82$\arcsecond$ & --- & ---           & $5.41  ^{+4.58} _{-4.22}$  & $<0.11$ & $0.76^{+0.53}_{-0.48}$                & $0.60 ^{+0.40}_{-0.04}$ & 24.68 & 24.31 & 22.92 & --- & --- & --- & --- & --- \\
X56 & 220.2962 & 39.9642 & 6.56$\arcsecond$ & --- & ---            & $6.32  ^{+3.80} _{-3.38}$  & $0.16^{+0.11}_{-0.09}$ & $0.10^{+0.25}_{-0.10}$ & $-0.30^{+0.22}_{-0.70}$ & --- & --- & --- & --- & --- & --- & --- & --- \\
X57 & 220.3071 & 40.1206 & 4.25$\arcsecond$ & --- & ---            & $5.81  ^{+3.11} _{-2.76}$  & $0.05^{+0.06}_{-0.04}$ & $0.27^{+0.20}_{-0.16}$ & $0.25 ^{+0.75}_{-0.18}$ & --- & --- & --- & --- & --- & --- & --- & --- \\
X58 & 220.3714 & 40.1209 & 4.01$\arcsecond$ & --- & ---            & $4.80  ^{+2.92} _{-2.56}$  & $<0.06$ & $0.42^{+0.25}_{-0.20}$                & $0.75 ^{+0.25}_{-0.04}$ & --- & --- & --- & --- & --- & --- & --- & --- \\
X59 & 220.3420 & 39.9598 & 6.15$\arcsecond$ & --- & ---            & $2.81  ^{+2.75} _{-2.42}$  & $0.07^{+0.08}_{-0.05}$ & $0.57^{+0.32}_{-0.28}$ & $-0.16^{+0.33}_{-0.84}$ & --- & --- & --- & --- & --- & --- & --- & --- \\
\enddata
\tablecomments{
Column (1) XID: Source designation; (2) R.A.: Right ascension (J2000) [deg]; (3) Dec.: Declination (J2000) [deg]; (4) $r_{95}$: 95\% encircled-energy PSF radius [arcsec]; (5) $z_{\rm spec}$: Spectroscopic redshift (SDSS/DESI/CO); (6) $z_{\rm phot}$: Photometric redshift computed by EAZY \citep{EAZY}; (7) Counts${\rm FB}$: Full-band (0.5-7 keV) net counts; (8) $f{\rm 0.5-2keV}$: Soft-band flux [erg cm$^{-2}$ s$^{-1}$]; (9) $f_{\rm 2-10keV}$: Hard-band flux [erg cm$^{-2}$ s$^{-1}$]; (10) HR: Hardness ratio $(H-S)/(H+S)$; (11)-(13) DESI $[grz]$ photometry [AB mag]; 
(14)-(17) WISE $[W1-W4]$ photometry [AB mag]; (18) Class: WISE classification \citep{2010AJ....140.1868W}. Catalog contains 59 X-ray sources after foreground exclusion. Fluxes computed with \texttt{srcflux}.
}
\end{deluxetable}
\end{longrotatetable}

\begin{deluxetable*}{ccccccc}
\tabletypesize{\small}
\tablecaption{Best-fit parameters for the simple power-law spectral model} \label{table:fit_result_1}
\tablehead{
    \colhead{(1)} & 
    \colhead{(2)} & 
    \colhead{(3)} & 
    \colhead{(4)} & 
    \colhead{(5)} & 
    \colhead{(6)} & 
    \colhead{(7)} \\  
    \colhead{$XID$} &
    \colhead{$K_{\rm 1keV}$} &
    \colhead{$\Gamma$} &
    \colhead{$f_{0.5-10~\rm keV}$} &
    \colhead{$L_{2-7~\rm keV}$} &
    \colhead{$L_{7-33~\rm keV}$} &
    \colhead{$\rm{C}/\nu$} \\
    \colhead{} & 
    \colhead{$10^{-5}$} & 
    \colhead{} & 
    \colhead{$\log{}~\rm erg~cm^{-2}~s^{-1}$} & 
    \colhead{$\log{}~\rm erg~s^{-1}$} & 
    \colhead{$\log{}~\rm erg~s^{-1}$} & 
    \colhead{}
}
\startdata
LAE10-X1 & $0.0019^{+0.45}_{}$  & $-0.68^{+2.12}_{}$  & $-13.92^{+0.50}_{-0.48}$ & $42.89^{+1.02}_{-1.07}$ & $44.70^{+0.51}_{-0.52}$ & $7.51/2$ \\
LAE-X2 & $0.20^{+22.87}_{-0.20}$ & $1.02^{+1.98}_{-2.17}$ & $-14.07^{+0.37}_{-0.29}$ & $43.78^{+0.95}_{-1.07}$ & $44.48^{+0.45}_{-0.40}$ & $2.92/3$ \\
LAE-X3 & $2.57^{+5.41}_{-1.75}$  & $1.61^{+0.58}_{-0.53}$ & $-13.58^{+0.14}_{-0.13}$ & $44.55^{+0.18}_{-0.19}$ & $44.89^{+0.21}_{-0.24}$ & $26.74/24$ \\
LAE11-X4 & $2.94^{+7.18}_{-2.12}$  & $2.01^{+0.69}_{-0.63}$ & $-13.90^{+0.14}_{-0.15}$ & $44.36^{+0.19}_{-0.21}$ & $44.44^{+0.27}_{-0.30}$ & $18.62/14$ \\
LAE-X5 & $2.30^{+44.32}_{-2.19}$ & $1.87^{+1.67}_{-1.48}$ & $-13.88^{+0.40}_{-0.28}$ & $44.33^{+0.45}_{-0.46}$ & $44.51^{+0.60}_{-0.64}$ & $1.64/2$ \\
LAE-X6 & $2.26^{+8.09}_{-1.78}$  & $2.06^{+0.85}_{-0.77}$ & $-14.06^{+0.17}_{-0.18}$ & $44.22^{+0.24}_{-0.26}$ & $44.28^{+0.32}_{-0.37}$ & $8.07/9$ \\
X7 & $21.62^{+28.43}_{-12.11}$ & $2.51^{+0.48}_{-0.44}$ & $-13.42^{+0.09}_{-0.09}$ & $44.94^{+0.13}_{-0.14}$ & $44.72^{+0.18}_{-0.20}$ & $29.66/35$ \\
\enddata
\tablecomments{Column (1) XID: Source designation; (2) $K_{\rm 1keV}$: Power-law normalization at 1 keV (10$^{-5}$ photons keV$^{-1}$ cm$^{-2}$ s$^{-1}$); (3) $\Gamma$: Photon index; (4) log $f_{\rm 0.5-10keV}$: Observed 0.5-10 keV flux (erg cm$^{-2}$ s$^{-1}$); (5) log $L_{\rm 2-7keV}$: Rest-frame 2-7 keV luminosity (erg s$^{-1}$); (6) log $L_{\rm 7-33keV}$: Rest-frame 7-33 keV luminosity (erg s$^{-1}$); (7) C/$\nu$: Cash statistic and degrees of freedom. All flux and luminosity values are corrected for Galactic absorption. For LAE-X2, we assume $z = 2.31$. 
}
\end{deluxetable*}

\begin{deluxetable*}{ccccccc}
\tabletypesize{\small}
\tablecaption{Best-fit parameters for the intrinsic absorbed power-law model ($\Gamma$ fixed to 1.8)} \label{table:fit_result_2}
\tablehead{
    \colhead{(1)} & 
    \colhead{(2)} & 
    \colhead{(3)} & 
    \colhead{(4)} & 
    \colhead{(5)} & 
    \colhead{(6)} & 
    \colhead{(7)} \\  
    \colhead{$XID$} &
    \colhead{$K_{\rm 1keV}$} &
    \colhead{$\rm N_H(z)$} &
    \colhead{$f_{0.5-10~\rm keV}$} &
    \colhead{$L_{2-7~\rm keV}$} &
    \colhead{$L_{7-33~\rm keV}$} &
    \colhead{$\rm{C}/\nu$} \\
    \colhead{} & 
    \colhead{$10^{-5}$} & 
    \colhead{$10^{22}~\rm cm^{-2}$} & 
    \colhead{$\log{}~\rm erg~cm^{-2}~s^{-1}$} & 
    \colhead{$\log{}~\rm erg~s^{-1}$} & 
    \colhead{$\log{}~\rm erg~s^{-1}$} & 
    \colhead{}
}
\startdata
LAE10-X1 & $4.35^{+9.13}_{-3.07}$ & $209.36^{+315.29}_{-169.01}$ & $-13.54^{+0.49}_{-0.53}$ & $44.65^{+0.49}_{-0.53}$ & $44.87^{+0.49}_{-0.53}$ & $6.51/2$ \\
LAE-X2 & $1.72^{+2.95}_{-1.01}$ & $<157.00$ & $-13.94^{+0.43}_{-0.38}$ & $44.26^{+0.43}_{-0.38}$ & $44.47^{+0.43}_{-0.38}$ & $3.16/3$ \\ 
LAE-X3 & $3.79^{+1.81}_{-0.87}$ & $<11.27$ & $-13.61^{+0.17}_{-0.11}$ & $44.0^{+0.17}_{-0.11}$ & $44.81^{+0.17}_{-0.11}$ & $27.07/24$ \\
LAE11-X4 & $1.97^{+0.70}_{-0.55}$ & $<4.90$ & $-13.88^{+0.13}_{-0.14}$ & $44.31^{+0.13}_{-0.14}$ & $44.52^{+0.13}_{-0.14}$ & $18.92/14$ \\
LAE-X5 & $2.02^{+2.74}_{-0.94}$ & $<29.23$ & $-13.87^{+0.37}_{-0.27}$ & $44.32^{+0.37}_{-0.27}$ & $44.53^{+0.37}_{-0.27}$ & $1.65/2$ \\
LAE-X6 & $1.39^{+0.66}_{-0.46}$ & $<7.50$ & $-14.04^{+0.17}_{-0.18}$ & $44.16^{+0.17}_{-0.18}$ & $44.37^{+0.17}_{-0.18}$ & $8.37/9$ \\
X7 & $5.71^{+1.19}_{-1.05}$ & $<2.52$ & $-13.42^{+0.08}_{-0.09}$ & $44.77^{+0.08}_{-0.09}$ & $44.98^{+0.08}_{-0.09}$ & $37.21/35$ \\
\enddata
\tablecomments{Column (1) XID: Source designation; (2) $K_{\rm 1keV}$: Power-law normalization at 1 keV [photons keV$^{-1}$ cm$^{-2}$ s$^{-1}$]; (3) $N_{\rm H}$: Intrinsic column density [10$^{22}$ cm$^{-2}$]; (4) log $f_{\rm 0.5-10keV}$: Observed 0.5-10 keV flux [erg cm$^{-2}$ s$^{-1}$]; (5) log $L_{\rm 2-7keV}$: Rest-frame 2-7 keV luminosity [erg s$^{-1}$]; (6) log $L_{\rm 7-33keV}$: Rest-frame 7-33 keV luminosity [erg s$^{-1}$]; (7) C-stat/dof: Cash statistic and degrees of freedom. All flux and luminosity values are corrected for both Galactic and intrinsic absorption. Luminosities assume $z = 2.31$ for LAE-X2.
}
\end{deluxetable*}

\begin{acknowledgments}

This work was supported by the Joint Research Foundation in Astronomy under the cooperative agreement between the National Science Foundation of China and the CAS (U1731104, U1731109), National Science Foundation of China (NSFC-11833007). L.M.D. also acknowledges the support from the Key Laboratory for Astronomical Observation and Technology of Guangzhou, the Astronomy Science and Technology Research Laboratory of the education department of Guangdong Province. 

\end{acknowledgments}

\bibliography{ms_v2}

\begin{thebibliography}{}
\expandafter\ifx\csname natexlab\endcsname\relax\def\natexlab#1{#1}\fi
\providecommand{\url}[1]{\href{#1}{#1}}
\providecommand{\dodoi}[1]{doi:~\href{http://doi.org/#1}{\nolinkurl{#1}}}
\providecommand{\doeprint}[1]{\href{http://ascl.net/#1}{\nolinkurl{http://ascl.net/#1}}}
\providecommand{\doarXiv}[1]{\href{https://arxiv.org/abs/#1}{\nolinkurl{https://arxiv.org/abs/#1}}}

\bibitem[{{Almeida} {et~al.}(2023){Almeida}, {Anderson}, {Argudo-Fern{\'a}ndez}, {Badenes}, {Barger}, {Barrera-Ballesteros}, {Bender}, {Benitez}, {Besser}, {Bird}, {Bizyaev}, {Blanton}, {Bochanski}, {Bovy}, {Brandt}, {Brownstein}, {Buchner}, {Bulbul}, {Burchett}, {Cano D{\'\i}az}, {Carlberg}, {Casey}, {Chandra}, {Cherinka}, {Chiappini}, {Coker}, {Comparat}, {Conroy}, {Contardo}, {Cortes}, {Covey}, {Crane}, {Cunha}, {Dabbieri}, {Davidson}, {Davis}, {de Andrade Queiroz}, {De Lee}, {M{\'e}ndez Delgado}, {Demasi}, {Di Mille}, {Donor}, {Dow}, {Dwelly}, {Eracleous}, {Eriksen}, {Fan}, {Farr}, {Frederick}, {Fries}, {Frinchaboy}, {G{\"a}nsicke}, {Ge}, {Gonz{\'a}lez {\'A}vila}, {Grabowski}, {Grier}, {Guiglion}, {Gupta}, {Hall}, {Hawkins}, {Hayes}, {Hermes}, {Hern{\'a}ndez-Garc{\'\i}a}, {Hogg}, {Holtzman}, {Ibarra-Medel}, {Ji}, {Jofre}, {Johnson}, {Jones}, {Kinemuchi}, {Kluge}, {Koekemoer}, {Kollmeier}, {Kounkel}, {Krishnarao}, {Krumpe}, {Lacerna}, {Lago}, {Laporte}, {Liu}, {Liu}, {Liu}, {Lopes}, {Macktoobian},
  {Majewski}, {Malanushenko}, {Maoz}, {Masseron}, {Masters}, {Matijevic}, {McBride}, {Medan}, {Merloni}, {Morrison}, {Myers}, {M{\'e}sz{\'a}ros}, {Negrete}, {Nidever}, {Nitschelm}, {Oravetz}, {Oravetz}, {Pan}, {Peng}, {Pinsonneault}, {Pogge}, {Qiu}, {Ramirez}, {Rix}, {Fern{\'a}ndez Rosso}, {Runnoe}, {Salvato}, {Sanchez}, {Santana}, {Saydjari}, {Sayres}, {Schlaufman}, {Schneider}, {Schwope}, {Serna}, {Shen}, {Sobeck}, {Song}, {Souto}, {Spoo}, {Stassun}, {Steinmetz}, {Straumit}, {Stringfellow}, {S{\'a}nchez-Gallego}, {Taghizadeh-Popp}, {Tayar}, {Thakar}, {Tissera}, {Tkachenko}, {Hernandez Toledo}, {Trakhtenbrot}, {Fern{\'a}ndez-Trincado}, {Troup}, {Trump}, {Tuttle}, {Ulloa}, {Vazquez-Mata}, {Vera Alfaro}, {Villanova}, {Wachter}, {Weijmans}, {Wheeler}, {Wilson}, {Wojno}, {Wolf}, {Xue}, {Ybarra}, {Zari}, \& {Zasowski}}]{2023ApJS..267...44A}
{Almeida}, A., {Anderson}, S.~F., {Argudo-Fern{\'a}ndez}, M., {et~al.} 2023, \apjs, 267, 44, \dodoi{10.3847/1538-4365/acda98}

\bibitem[{{Arnaud}(1996)}]{XSPEC}
{Arnaud}, K.~A. 1996, in Astronomical Society of the Pacific Conference Series, Vol. 101, Astronomical Data Analysis Software and Systems V, ed. G.~H. {Jacoby} \& J.~{Barnes}, 17

\bibitem[{{Arrigoni Battaia} {et~al.}(2018){Arrigoni Battaia}, {Chen}, {Fumagalli}, {Cai}, {Calistro Rivera}, {Xu}, {Smail}, {Prochaska}, {Yang}, \& {De Breuck}}]{2018A&A...620A.202A}
{Arrigoni Battaia}, F., {Chen}, C.-C., {Fumagalli}, M., {et~al.} 2018, \aap, 620, A202, \dodoi{10.1051/0004-6361/201834195}

\bibitem[{{Bohlin} {et~al.}(1978){Bohlin}, {Savage}, \& {Drake}}]{1978ApJ...224..132B}
{Bohlin}, R.~C., {Savage}, B.~D., \& {Drake}, J.~F. 1978, \apj, 224, 132, \dodoi{10.1086/156357}

\bibitem[{{Brammer} {et~al.}(2008){Brammer}, {van Dokkum}, \& {Coppi}}]{EAZY}
{Brammer}, G.~B., {van Dokkum}, P.~G., \& {Coppi}, P. 2008, \apj, 686, 1503, \dodoi{10.1086/591786}

\bibitem[{{Cai} {et~al.}(2016){Cai}, {Fan}, {Peirani}, {Bian}, {Frye}, {McGreer}, {Prochaska}, {Lau}, {Tejos}, {Ho}, \& {Schneider}}]{2016ApJ...833..135C}
{Cai}, Z., {Fan}, X., {Peirani}, S., {et~al.} 2016, \apj, 833, 135, \dodoi{10.3847/1538-4357/833/2/135}

\bibitem[{{Cai} {et~al.}(2017{\natexlab{a}}){Cai}, {Fan}, {Bian}, {Zabludoff}, {Yang}, {Prochaska}, {McGreer}, {Zheng}, {Kashikawa}, {Wang}, {Frye}, {Green}, \& {Jiang}}]{2017ApJ...839..131C}
{Cai}, Z., {Fan}, X., {Bian}, F., {et~al.} 2017{\natexlab{a}}, \apj, 839, 131, \dodoi{10.3847/1538-4357/aa6a1a}

\bibitem[{{Cai} {et~al.}(2017{\natexlab{b}}){Cai}, {Fan}, {Yang}, {Bian}, {Prochaska}, {Zabludoff}, {McGreer}, {Zheng}, {Green}, {Cantalupo}, {Frye}, {Hamden}, {Jiang}, {Kashikawa}, \& {Wang}}]{2017ApJ...837...71C}
{Cai}, Z., {Fan}, X., {Yang}, Y., {et~al.} 2017{\natexlab{b}}, \apj, 837, 71, \dodoi{10.3847/1538-4357/aa5d14}

\bibitem[{{Cai} {et~al.}(2018){Cai}, {Hamden}, {Matuszewski}, {Prochaska}, {Li}, {Cantalupo}, {Arrigoni Battaia}, {Martin}, {Neill}, {O'Sullivan}, {Wang}, {Moore}, \& {Morrissey}}]{2018ApJ...861L...3C}
{Cai}, Z., {Hamden}, E., {Matuszewski}, M., {et~al.} 2018, \apjl, 861, L3, \dodoi{10.3847/2041-8213/aacce6}

\bibitem[{{Cantalupo} {et~al.}(2014){Cantalupo}, {Arrigoni-Battaia}, {Prochaska}, {Hennawi}, \& {Madau}}]{2014Natur.506...63C}
{Cantalupo}, S., {Arrigoni-Battaia}, F., {Prochaska}, J.~X., {Hennawi}, J.~F., \& {Madau}, P. 2014, \nat, 506, 63, \dodoi{10.1038/nature12898}

\bibitem[{{Carilli} {et~al.}(1997){Carilli}, {R{\"o}ttgering}, {van Ojik}, {Miley}, \& {van Breugel}}]{1997ApJS..109....1C}
{Carilli}, C.~L., {R{\"o}ttgering}, H.~J.~A., {van Ojik}, R., {Miley}, G.~K., \& {van Breugel}, W.~J.~M. 1997, \apjs, 109, 1, \dodoi{10.1086/312973}

\bibitem[{{Cash}(1979)}]{1979ApJ...228..939C}
{Cash}, W. 1979, \apj, 228, 939, \dodoi{10.1086/156922}

\bibitem[{{Chapman} {et~al.}(2004){Chapman}, {Scott}, {Windhorst}, {Frayer}, {Borys}, {Lewis}, \& {Ivison}}]{Chapman2004ApJ}
{Chapman}, S.~C., {Scott}, D., {Windhorst}, R.~A., {et~al.} 2004, \apj, 606, 85, \dodoi{10.1086/382778}

\bibitem[{{Chiang} {et~al.}(2014){Chiang}, {Overzier}, \& {Gebhardt}}]{2014ApJ...782L...3C}
{Chiang}, Y.-K., {Overzier}, R., \& {Gebhardt}, K. 2014, \apjl, 782, L3, \dodoi{10.1088/2041-8205/782/1/L3}

\bibitem[{{DESI Collaboration} {et~al.}(2025){DESI Collaboration}, {Abdul-Karim}, {Adame}, {Aguado}, {Aguilar}, {Ahlen}, {Alam}, {Aldering}, {Alexander}, {Alfarsy}, {Allen}, {Allende Prieto}, {Alves}, {Anand}, {Andrade}, {Armengaud}, {Avila}, {Aviles}, {Awan}, {Bailey}, {Baleato Lizancos}, {Ballester}, {Bault}, {Bautista}, {BenZvi}, {Beraldo e Silva}, {Bermejo-Climent}, {Beutler}, {Bianchi}, {Blake}, {Blum}, {Bolton}, {Bonici}, {Brieden}, {Brodzeller}, {Brooks}, {Buckley-Geer}, {Burtin}, {Canning}, {Carnero Rosell}, {Carr}, {Carrilho}, {Casas}, {Castander}, {Cereskaite}, {Cervantes-Cota}, {Chaussidon}, {Chaves-Montero}, {Chen}, {Chen}, {Claybaugh}, {Cole}, {Cooper}, {Cousinou}, {Cuceu}, {Davis}, {Dawson}, {de Belsunce}, {de la Cruz}, {de la Macorra}, {de Mattia}, {Deiosso}, {Della Costa}, {Demina}, {Demirbozan}, {DeRose}, {Dey}, {Dey}, {Ding}, {Ding}, {Doel}, {Douglass}, {Dowicz}, {Ebina}, {Edelstein}, {Eisenstein}, {Elbers}, {Emas}, {Escoffier}, {Fagrelius}, {Fan}, {Fanning}, {Fawcett},
  {Fern\textbackslash'andez-Garc\textbackslash'ia}, {Ferraro}, {Findlay}, {Font-Ribera}, {Forero-Romero}, {Forero-S\textbackslash'anchez}, {Frenk}, {G\textbackslash''ansicke}, {Galbany}, {Garc\textbackslash'ia-Bellido}, {Garcia-Quintero}, {Garrison}, {Gazta\textbackslash\raisebox{-0.5ex}\textasciitilde naga}, {Gil-Mar\textbackslash'in}, {Gnedin}, {Gontcho}, {Gonzalez-Morales}, {Gonzalez-Perez}, {Gordon}, {Graur}, {Green}, {Gruen}, {Gsponer}, {Guandalin}, {Gutierrez}, {Guy}, {Hahn}, {Han}, {Han}, {He}, {Herrera-Alcantar}, {Honscheid}, {Hou}, {Howlett}, {Huterer}, {Ir\textbackslash v\{s\}i\textbackslash v\{c\}}, {Ishak}, {Jacques}, {Jimenez}, {Jing}, {Joachimi}, {Joudaki}, {Joyce}, {Jullo}, {Juneau}, {Kara\textbackslash c\{c\}ayl\{\textbackslash i\}}, {Karim}, {Kehoe}, {Kent}, {Khederlarian}, {Kirkby}, {Kisner}, {Kitaura}, {Kizhuprakkat}, {Kong}, {Koposov}, {Kremin}, {Krolewski}, {Lahav}, {Lai}, {Lamman}, {Lan}, {Landriau}, {Lang}, {Lange}, {Lasker}, {Le Goff}, {Le Guillou}, {Leauthaud}, {Levi}, {Li}, {Li},
  {Lodha}, {Lokken}, {Luo}, {Magneville}, {Manera}, {Manser}, {Margala}, {Martini}, {Maus}, {McCullough}, {McDonald}, {Medina}, {Medina-Varela}, {Meisner}, {Mena-Fern\textbackslash'andez}, {Menegas}, {Mezcua}, {Miquel}, {Montero-Camacho}, {Moon}, {Moustakas}, {Mu\textbackslash\raisebox{-0.5ex}\textasciitilde noz-Guti\textbackslash'errez}, {Mu\textbackslash\raisebox{-0.5ex}\textasciitilde noz-Santos}, {Myers}, {Myles}, {Nadathur}, {Najita}, {Napolitano}, {Newman}, {Nikakhtar}, {Nikutta}, {Niz}, {Noriega}, {Padmanabhan}, {Paillas}, {Palanque-Delabrouille}, {Palmese}, {Pan}, {Pan}, {Parkinson}, {Peacock}, {Percival}, {P\textbackslash'erez-Fern\textbackslash'andez}, {P\textbackslash'erez-R\textbackslash`afols}, \& {Peterson}}]{2025arXiv250314745D}
{DESI Collaboration}, {Abdul-Karim}, M., {Adame}, A.~G., {et~al.} 2025, arXiv e-prints, arXiv:2503.14745, \dodoi{10.48550/arXiv.2503.14745}

\bibitem[{{Dey} {et~al.}(2019){Dey}, {Schlegel}, {Lang}, {Blum}, {Burleigh}, {Fan}, {Findlay}, {Finkbeiner}, {Herrera}, {Juneau}, {Landriau}, {Levi}, {McGreer}, {Meisner}, {Myers}, {Moustakas}, {Nugent}, {Patej}, {Schlafly}, {Walker}, {Valdes}, {Weaver}, {Y{\`e}che}, {Zou}, {Zhou}, {Abareshi}, {Abbott}, {Abolfathi}, {Aguilera}, {Alam}, {Allen}, {Alvarez}, {Annis}, {Ansarinejad}, {Aubert}, {Beechert}, {Bell}, {BenZvi}, {Beutler}, {Bielby}, {Bolton}, {Brice{\~n}o}, {Buckley-Geer}, {Butler}, {Calamida}, {Carlberg}, {Carter}, {Casas}, {Castander}, {Choi}, {Comparat}, {Cukanovaite}, {Delubac}, {DeVries}, {Dey}, {Dhungana}, {Dickinson}, {Ding}, {Donaldson}, {Duan}, {Duckworth}, {Eftekharzadeh}, {Eisenstein}, {Etourneau}, {Fagrelius}, {Farihi}, {Fitzpatrick}, {Font-Ribera}, {Fulmer}, {G{\"a}nsicke}, {Gaztanaga}, {George}, {Gerdes}, {Gontcho}, {Gorgoni}, {Green}, {Guy}, {Harmer}, {Hernandez}, {Honscheid}, {Huang}, {James}, {Jannuzi}, {Jiang}, {Joyce}, {Karcher}, {Karkar}, {Kehoe}, {Kneib}, {Kueter-Young}, {Lan},
  {Lauer}, {Le Guillou}, {Le Van Suu}, {Lee}, {Lesser}, {Perreault Levasseur}, {Li}, {Mann}, {Marshall}, {Mart{\'\i}nez-V{\'a}zquez}, {Martini}, {du Mas des Bourboux}, {McManus}, {Meier}, {M{\'e}nard}, {Metcalfe}, {Mu{\~n}oz-Guti{\'e}rrez}, {Najita}, {Napier}, {Narayan}, {Newman}, {Nie}, {Nord}, {Norman}, {Olsen}, {Paat}, {Palanque-Delabrouille}, {Peng}, {Poppett}, {Poremba}, {Prakash}, {Rabinowitz}, {Raichoor}, {Rezaie}, {Robertson}, {Roe}, {Ross}, {Ross}, {Rudnick}, {Safonova}, {Saha}, {S{\'a}nchez}, {Savary}, {Schweiker}, {Scott}, {Seo}, {Shan}, {Silva}, {Slepian}, {Soto}, {Sprayberry}, {Staten}, {Stillman}, {Stupak}, {Summers}, {Sien Tie}, {Tirado}, {Vargas-Maga{\~n}a}, {Vivas}, {Wechsler}, {Williams}, {Yang}, {Yang}, {Yapici}, {Zaritsky}, {Zenteno}, {Zhang}, {Zhang}, {Zhou}, \& {Zhou}}]{2019AJ....157..168D}
{Dey}, A., {Schlegel}, D.~J., {Lang}, D., {et~al.} 2019, \aj, 157, 168, \dodoi{10.3847/1538-3881/ab089d}

\bibitem[{Digby-North {et~al.}(2010)Digby-North, Nandra, Laird, Steidel, Georgakakis, Bogosavljević, Erb, Shapley, Reddy, \& Aird}]{10.1111/j.1365-2966.2010.16977.x}
Digby-North, J.~A., Nandra, K., Laird, E.~S., {et~al.} 2010, Monthly Notices of the Royal Astronomical Society, 407, 846, \dodoi{10.1111/j.1365-2966.2010.16977.x}

\bibitem[{{Dressler}(1980)}]{1980ApJ...236..351D}
{Dressler}, A. 1980, \apj, 236, 351, \dodoi{10.1086/157753}

\bibitem[{{Elbaz} {et~al.}(2007){Elbaz}, {Daddi}, {Le Borgne}, {Dickinson}, {Alexander}, {Chary}, {Starck}, {Brandt}, {Kitzbichler}, {MacDonald}, {Nonino}, {Popesso}, {Stern}, \& {Vanzella}}]{2007A&A...468...33E}
{Elbaz}, D., {Daddi}, E., {Le Borgne}, D., {et~al.} 2007, \aap, 468, 33, \dodoi{10.1051/0004-6361:20077525}

\bibitem[{{Emonts} {et~al.}(2019){Emonts}, {Cai}, {Prochaska}, {Li}, \& {Lehnert}}]{2019ApJ...887...86E}
{Emonts}, B. H.~C., {Cai}, Z., {Prochaska}, J.~X., {Li}, Q., \& {Lehnert}, M.~D. 2019, \apj, 887, 86, \dodoi{10.3847/1538-4357/ab45f4}

\bibitem[{Evans {et~al.}(2024)Evans, Evans, Martínez-Galarza, Miller, Primini, Azadi, Burke, Civano, D’Abrusco, Fabbiano, Graessle, Grier, Houck, Lauer, McCollough, Nowak, Plummer, Rots, Siemiginowska, \& Tibbetts}]{Evans2024}
Evans, I.~N., Evans, J.~D., Martínez-Galarza, J.~R., {et~al.} 2024, The Astrophysical Journal Supplement Series, 274, 22, \dodoi{10.3847/1538-4365/ad6319}

\bibitem[{{Freeman} {et~al.}(2002){Freeman}, {Kashyap}, {Rosner}, \& {Lamb}}]{WAVDETECT}
{Freeman}, P.~E., {Kashyap}, V., {Rosner}, R., \& {Lamb}, D.~Q. 2002, \apjs, 138, 185, \dodoi{10.1086/324017}

\bibitem[{{Fruscione} {et~al.}(2006){Fruscione}, {McDowell}, {Allen}, {Brickhouse}, {Burke}, {Davis}, {Durham}, {Elvis}, {Galle}, {Harris}, {Huenemoerder}, {Houck}, {Ishibashi}, {Karovska}, {Nicastro}, {Noble}, {Nowak}, {Primini}, {Siemiginowska}, {Smith}, \& {Wise}}]{CIAO}
{Fruscione}, A., {McDowell}, J.~C., {Allen}, G.~E., {et~al.} 2006, in Society of Photo-Optical Instrumentation Engineers (SPIE) Conference Series, Vol. 6270, Society of Photo-Optical Instrumentation Engineers (SPIE) Conference Series, ed. D.~R. {Silva} \& R.~E. {Doxsey}, 62701V, \dodoi{10.1117/12.671760}

\bibitem[{{Garmire} {et~al.}(2003){Garmire}, {Bautz}, {Ford}, {Nousek}, \& {Ricker}}]{ACIS}
{Garmire}, G.~P., {Bautz}, M.~W., {Ford}, P.~G., {Nousek}, J.~A., \& {Ricker}, George~R., J. 2003, in Society of Photo-Optical Instrumentation Engineers (SPIE) Conference Series, Vol. 4851, X-Ray and Gamma-Ray Telescopes and Instruments for Astronomy., ed. J.~E. {Truemper} \& H.~D. {Tananbaum}, 28--44, \dodoi{10.1117/12.461599}

\bibitem[{{Gehrels}(1986)}]{1986ApJ...303..336G}
{Gehrels}, N. 1986, \apj, 303, 336, \dodoi{10.1086/164079}

\bibitem[{{Gilli} {et~al.}(2007){Gilli}, {Comastri}, \& {Hasinger}}]{2007A&A...463...79G}
{Gilli}, R., {Comastri}, A., \& {Hasinger}, G. 2007, \aap, 463, 79, \dodoi{10.1051/0004-6361:20066334}

\bibitem[{{Gilli} {et~al.}(2019){Gilli}, {Mignoli}, {Peca}, {Nanni}, {Prandoni}, {Liuzzo}, {D'Amato}, {Brusa}, {Calura}, {Caminha}, {Chiaberge}, {Comastri}, {Cucciati}, {Cusano}, {Grandi}, {Decarli}, {Lanzuisi}, {Mannucci}, {Pinna}, {Tozzi}, {Vanzella}, {Vignali}, {Vito}, {Balmaverde}, {Citro}, {Cappelluti}, {Zamorani}, \& {Norman}}]{2019A&A...632A..26G}
{Gilli}, R., {Mignoli}, M., {Peca}, A., {et~al.} 2019, \aap, 632, A26, \dodoi{10.1051/0004-6361/201936121}

\bibitem[{{Goto} {et~al.}(2003){Goto}, {Yamauchi}, {Fujita}, {Okamura}, {Sekiguchi}, {Smail}, {Bernardi}, \& {Gomez}}]{2003MNRAS.346..601G}
{Goto}, T., {Yamauchi}, C., {Fujita}, Y., {et~al.} 2003, \mnras, 346, 601, \dodoi{10.1046/j.1365-2966.2003.07114.x}

\bibitem[{{Gr{\"u}tzbauch} {et~al.}(2011){Gr{\"u}tzbauch}, {Conselice}, {Bauer}, {Bluck}, {Chuter}, {Buitrago}, {Mortlock}, {Weinzirl}, \& {Jogee}}]{2011MNRAS.418..938G}
{Gr{\"u}tzbauch}, R., {Conselice}, C.~J., {Bauer}, A.~E., {et~al.} 2011, \mnras, 418, 938, \dodoi{10.1111/j.1365-2966.2011.19559.x}

\bibitem[{{Hashiguchi} {et~al.}(2023){Hashiguchi}, {Toba}, {Ota}, {Oguri}, {Okabe}, {Ueda}, {Imanishi}, {Yamada}, {Goto}, {Koyama}, {Lee}, {Mitsuishi}, {Nagao}, {Nishizawa}, {Noboriguchi}, {Oogi}, {Sakuta}, {Schramm}, {Shibata}, {Terashima}, {Yamashita}, {Yanagawa}, \& {Yoshimoto}}]{2023PASJ...75.1246H}
{Hashiguchi}, A., {Toba}, Y., {Ota}, N., {et~al.} 2023, \pasj, 75, 1246, \dodoi{10.1093/pasj/psad066}

\bibitem[{{Hennawi} {et~al.}(2015){Hennawi}, {Prochaska}, {Cantalupo}, \& {Arrigoni-Battaia}}]{2015Sci...348..779H}
{Hennawi}, J.~F., {Prochaska}, J.~X., {Cantalupo}, S., \& {Arrigoni-Battaia}, F. 2015, Science, 348, 779, \dodoi{10.1126/science.aaa5397}

\bibitem[{{HI4PI Collaboration} {et~al.}(2016){HI4PI Collaboration}, {Ben Bekhti}, {Fl{\"o}er}, {Keller}, {Kerp}, {Lenz}, {Winkel}, {Bailin}, {Calabretta}, {Dedes}, {Ford}, {Gibson}, {Haud}, {Janowiecki}, {Kalberla}, {Lockman}, {McClure-Griffiths}, {Murphy}, {Nakanishi}, {Pisano}, \& {Staveley-Smith}}]{2016A&A...594A.116H}
{HI4PI Collaboration}, {Ben Bekhti}, N., {Fl{\"o}er}, L., {et~al.} 2016, \aap, 594, A116, \dodoi{10.1051/0004-6361/201629178}

\bibitem[{{Hu} {et~al.}(1996){Hu}, {McMahon}, \& {Egami}}]{Hu1996ApJ}
{Hu}, E.~M., {McMahon}, R.~G., \& {Egami}, E. 1996, \apjl, 459, L53, \dodoi{10.1086/309958}

\bibitem[{{Kaastra}(2017)}]{2017A&A...605A..51K}
{Kaastra}, J.~S. 2017, \aap, 605, A51, \dodoi{10.1051/0004-6361/201629319}

\bibitem[{{Koyama} {et~al.}(2013){Koyama}, {Kodama}, {Tadaki}, {Hayashi}, {Tanaka}, {Smail}, {Tanaka}, \& {Kurk}}]{2013MNRAS.428.1551K}
{Koyama}, Y., {Kodama}, T., {Tadaki}, K.-i., {et~al.} 2013, \mnras, 428, 1551, \dodoi{10.1093/mnras/sts133}

\bibitem[{{Lehmer} {et~al.}(2009){Lehmer}, {Alexander}, {Geach}, {Smail}, {Basu-Zych}, {Bauer}, {Chapman}, {Matsuda}, {Scharf}, {Volonteri}, \& {Yamada}}]{2009ApJ...691..687L}
{Lehmer}, B.~D., {Alexander}, D.~M., {Geach}, J.~E., {et~al.} 2009, \apj, 691, 687, \dodoi{10.1088/0004-637X/691/1/687}

\bibitem[{{Lehmer} {et~al.}(2013){Lehmer}, {Lucy}, {Alexander}, {Best}, {Geach}, {Harrison}, {Hornschemeier}, {Matsuda}, {Mullaney}, {Smail}, {Sobral}, \& {Swinbank}}]{2013ApJ...765...87L}
{Lehmer}, B.~D., {Lucy}, A.~B., {Alexander}, D.~M., {et~al.} 2013, \apj, 765, 87, \dodoi{10.1088/0004-637X/765/2/87}

\bibitem[{{Li} {et~al.}(2021){Li}, {Emonts}, {Cai}, {Prochaska}, {Yoon}, {Lehnert}, {Zhang}, {Wu}, {Li}, {Li}, {Lacy}, \& {Villar-Mart{\'\i}n}}]{2021ApJ...922L..29L}
{Li}, J., {Emonts}, B. H.~C., {Cai}, Z., {et~al.} 2021, \apjl, 922, L29, \dodoi{10.3847/2041-8213/ac390d}

\bibitem[{{Luo} {et~al.}(2017){Luo}, {Brandt}, {Xue}, {Lehmer}, {Alexander}, {Bauer}, {Vito}, {Yang}, {Basu-Zych}, {Comastri}, {Gilli}, {Gu}, {Hornschemeier}, {Koekemoer}, {Liu}, {Mainieri}, {Paolillo}, {Ranalli}, {Rosati}, {Schneider}, {Shemmer}, {Smail}, {Sun}, {Tozzi}, {Vignali}, \& {Wang}}]{2017ApJS..228....2L}
{Luo}, B., {Brandt}, W.~N., {Xue}, Y.~Q., {et~al.} 2017, \apjs, 228, 2, \dodoi{10.3847/1538-4365/228/1/2}

\bibitem[{{Macuga} {et~al.}(2019){Macuga}, {Martini}, {Miller}, {Brodwin}, {Hayashi}, {Kodama}, {Koyama}, {Overzier}, {Shimakawa}, {Tadaki}, \& {Tanaka}}]{2019ApJ...874...54M}
{Macuga}, M., {Martini}, P., {Miller}, E.~D., {et~al.} 2019, \apj, 874, 54, \dodoi{10.3847/1538-4357/ab0746}

\bibitem[{{Martini} {et~al.}(2013){Martini}, {Miller}, {Brodwin}, {Stanford}, {Gonzalez}, {Bautz}, {Hickox}, {Stern}, {Eisenhardt}, {Galametz}, {Norman}, {Jannuzi}, {Dey}, {Murray}, {Jones}, \& {Brown}}]{2013ApJ...768....1M}
{Martini}, P., {Miller}, E.~D., {Brodwin}, M., {et~al.} 2013, \apj, 768, 1, \dodoi{10.1088/0004-637X/768/1/1}

\bibitem[{{Park} {et~al.}(2006){Park}, {Kashyap}, {Siemiginowska}, {van Dyk}, {Zezas}, {Heinke}, \& {Wargelin}}]{BEHR}
{Park}, T., {Kashyap}, V.~L., {Siemiginowska}, A., {et~al.} 2006, \apj, 652, 610, \dodoi{10.1086/507406}

\bibitem[{{Pensabene} {et~al.}(2024){Pensabene}, {Cantalupo}, {Cicone}, {Decarli}, {Galbiati}, {Ginolfi}, {de Beer}, {Fossati}, {Fumagalli}, {Lazeyras}, {Pezzulli}, {Travascio}, {Wang}, {Matthee}, \& {Maseda}}]{Pensabene2024AA}
{Pensabene}, A., {Cantalupo}, S., {Cicone}, C., {et~al.} 2024, \aap, 684, A119, \dodoi{10.1051/0004-6361/202348659}

\bibitem[{{Polletta} {et~al.}(2021){Polletta}, {Soucail}, {Dole}, {Lehnert}, {Pointecouteau}, {Vietri}, {Scodeggio}, {Montier}, {Koyama}, {Lagache}, {Frye}, {Cusano}, \& {Fumana}}]{2021A&A...654A.121P}
{Polletta}, M., {Soucail}, G., {Dole}, H., {et~al.} 2021, \aap, 654, A121, \dodoi{10.1051/0004-6361/202140612}

\bibitem[{{Shi} {et~al.}(2021){Shi}, {Cai}, {Fan}, {Zheng}, {Huang}, \& {Xu}}]{2021ApJ...915...32S}
{Shi}, D.~D., {Cai}, Z., {Fan}, X., {et~al.} 2021, \apj, 915, 32, \dodoi{10.3847/1538-4357/abfec0}

\bibitem[{Shimakawa {et~al.}(2018)Shimakawa, Koyama, Röttgering, Kodama, Hayashi, Hatch, Dannerbauer, Tanaka, Tadaki, Suzuki, Fukagawa, Cai, \& Kurk}]{10.1093/mnras/sty2618}
Shimakawa, R., Koyama, Y., Röttgering, H. J.~A., {et~al.} 2018, Monthly Notices of the Royal Astronomical Society, 481, 5630, \dodoi{10.1093/mnras/sty2618}

\bibitem[{{Smith} {et~al.}(2009){Smith}, {Lucey}, {Hudson}, {Allanson}, {Bridges}, {Hornschemeier}, {Marzke}, \& {Miller}}]{2009MNRAS.392.1265S}
{Smith}, R.~J., {Lucey}, J.~R., {Hudson}, M.~J., {et~al.} 2009, \mnras, 392, 1265, \dodoi{10.1111/j.1365-2966.2008.14180.x}

\bibitem[{{Steidel} {et~al.}(1998){Steidel}, {Adelberger}, {Dickinson}, {Giavalisco}, {Pettini}, \& {Kellogg}}]{Steidel1998}
{Steidel}, C.~C., {Adelberger}, K.~L., {Dickinson}, M., {et~al.} 1998, \apj, 492, 428, \dodoi{10.1086/305073}

\bibitem[{{Steidel} {et~al.}(2005){Steidel}, {Adelberger}, {Shapley}, {Erb}, {Reddy}, \& {Pettini}}]{Steidel2005}
{Steidel}, C.~C., {Adelberger}, K.~L., {Shapley}, A.~E., {et~al.} 2005, \apj, 626, 44, \dodoi{10.1086/429989}

\bibitem[{{Steidel} {et~al.}(2000){Steidel}, {Adelberger}, {Shapley}, {Pettini}, {Dickinson}, \& {Giavalisco}}]{Steidel2000}
---. 2000, \apj, 532, 170, \dodoi{10.1086/308568}

\bibitem[{{Toshikawa} {et~al.}(2016){Toshikawa}, {Kashikawa}, {Overzier}, {Malkan}, {Furusawa}, {Ishikawa}, {Onoue}, {Ota}, {Tanaka}, {Niino}, \& {Uchiyama}}]{2016ApJ...826..114T}
{Toshikawa}, J., {Kashikawa}, N., {Overzier}, R., {et~al.} 2016, \apj, 826, 114, \dodoi{10.3847/0004-637X/826/2/114}

\bibitem[{{Tozzi} {et~al.}(2022){Tozzi}, {Pentericci}, {Gilli}, {Pannella}, {Fiore}, {Miley}, {Nonino}, {R{\"o}ttgering}, {Strazzullo}, {Anderson}, {Borgani}, {Calabr{\`o}}, {Carilli}, {Dannerbauer}, {Di Mascolo}, {Feruglio}, {Gobat}, {Jin}, {Liu}, {Mroczkowski}, {Norman}, {Rasia}, {Rosati}, \& {Saro}}]{2022A&A...662A..54T}
{Tozzi}, P., {Pentericci}, L., {Gilli}, R., {et~al.} 2022, \aap, 662, A54, \dodoi{10.1051/0004-6361/202142333}

\bibitem[{{Traina} {et~al.}(2025){Traina}, {Vito}, {Arrigoni-Battaia}, {Chen}, {Vignali}, {Prochaska}, {Cantalupo}, {Pensabene}, {Tozzi}, {Travascio}, {Gilli}, {Isla Llave}, {Marchesi}, \& {Mazzolari}}]{Traina2025arXiv}
{Traina}, A., {Vito}, F., {Arrigoni-Battaia}, F., {et~al.} 2025, arXiv e-prints, arXiv:2507.03078, \dodoi{10.48550/arXiv.2507.03078}

\bibitem[{{Tran} {et~al.}(2010){Tran}, {Papovich}, {Saintonge}, {Brodwin}, {Dunlop}, {Farrah}, {Finkelstein}, {Finkelstein}, {Lotz}, {McLure}, {Momcheva}, \& {Willmer}}]{2010ApJ...719L.126T}
{Tran}, K.-V.~H., {Papovich}, C., {Saintonge}, A., {et~al.} 2010, \apjl, 719, L126, \dodoi{10.1088/2041-8205/719/2/L126}

\bibitem[{{Travascio} {et~al.}(2025){Travascio}, {Cantalupo}, {Tozzi}, {Vito}, {Pezzulli}, {Paggi}, {Elvis}, {Fabbiano}, {Fiore}, {Fossati}, {Fresco}, {Fumagalli}, {Galbiati}, {Lazeyras}, {Ledos}, {Pannella}, {Pensabene}, {Quadri}, \& {Wang}}]{Travascio2025AA}
{Travascio}, A., {Cantalupo}, S., {Tozzi}, P., {et~al.} 2025, \aap, 694, A165, \dodoi{10.1051/0004-6361/202452179}

\bibitem[{{Venemans} {et~al.}(2007){Venemans}, {R{\"o}ttgering}, {Miley}, {van Breugel}, {de Breuck}, {Kurk}, {Pentericci}, {Stanford}, {Overzier}, {Croft}, \& {Ford}}]{Venemans2007AA}
{Venemans}, B.~P., {R{\"o}ttgering}, H.~J.~A., {Miley}, G.~K., {et~al.} 2007, \aap, 461, 823, \dodoi{10.1051/0004-6361:20053941}

\bibitem[{{Vito} {et~al.}(2020){Vito}, {Brandt}, {Lehmer}, {Vignali}, {Zou}, {Bauer}, {Bremer}, {Gilli}, {Ivison}, \& {Spingola}}]{2020A&A...642A.149V}
{Vito}, F., {Brandt}, W.~N., {Lehmer}, B.~D., {et~al.} 2020, \aap, 642, A149, \dodoi{10.1051/0004-6361/202038848}

\bibitem[{{Vito} {et~al.}(2024){Vito}, {Brandt}, {Comastri}, {Gilli}, {Ivison}, {Lanzuisi}, {Lehmer}, {Lopez}, {Tozzi}, \& {Vignali}}]{2024A&A...689A.130V}
{Vito}, F., {Brandt}, W.~N., {Comastri}, A., {et~al.} 2024, \aap, 689, A130, \dodoi{10.1051/0004-6361/202450225}

\bibitem[{{Wright} {et~al.}(2010){Wright}, {Eisenhardt}, {Mainzer}, {Ressler}, {Cutri}, {Jarrett}, {Kirkpatrick}, {Padgett}, {McMillan}, {Skrutskie}, {Stanford}, {Cohen}, {Walker}, {Mather}, {Leisawitz}, {Gautier}, {McLean}, {Benford}, {Lonsdale}, {Blain}, {Mendez}, {Irace}, {Duval}, {Liu}, {Royer}, {Heinrichsen}, {Howard}, {Shannon}, {Kendall}, {Walsh}, {Larsen}, {Cardon}, {Schick}, {Schwalm}, {Abid}, {Fabinsky}, {Naes}, \& {Tsai}}]{2010AJ....140.1868W}
{Wright}, E.~L., {Eisenhardt}, P. R.~M., {Mainzer}, A.~K., {et~al.} 2010, \aj, 140, 1868, \dodoi{10.1088/0004-6256/140/6/1868}

\bibitem[{{Wu} \& {Shen}(2022)}]{2022ApJS..263...42W}
{Wu}, Q., \& {Shen}, Y. 2022, \apjs, 263, 42, \dodoi{10.3847/1538-4365/ac9ead}

\bibitem[{{Yang} {et~al.}(2009){Yang}, {Zabludoff}, {Tremonti}, {Eisenstein}, \& {Dav{\'e}}}]{YangYujin2009ApJ}
{Yang}, Y., {Zabludoff}, A., {Tremonti}, C., {Eisenstein}, D., \& {Dav{\'e}}, R. 2009, \apj, 693, 1579, \dodoi{10.1088/0004-637X/693/2/1579}

\bibitem[{{Zhang} {et~al.}(2023){Zhang}, {Cai}, {Xu}, {Shimakawa}, {Arrigoni Battaia}, {Prochaska}, {Cen}, {Zheng}, {Wu}, {Li}, {Dou}, {Wu}, {Zabludoff}, {Fan}, {Ai}, {Golden-Marx}, {Li}, {Lu}, {Ma}, {Wang}, {Wang}, \& {Yuan}}]{2023Sci...380..494Z}
{Zhang}, S., {Cai}, Z., {Xu}, D., {et~al.} 2023, Science, 380, 494, \dodoi{10.1126/science.abj9192}

\end{thebibliography}
\bibliographystyle{aasjournal}

\end{document}